\documentclass[manuscript,screen]{acmart}

\AtBeginDocument{%
  \providecommand\BibTeX{{%
    \normalfont B\kern-0.5em{\scshape i\kern-0.25em b}\kern-0.8em\TeX}}}

\acmJournal{CSUR}

\usepackage[bottom]{footmisc}
\usepackage{multirow, makecell}
\usepackage{tabularx}
\usepackage{graphicx}
\DeclareGraphicsExtensions{.pdf}
\usepackage{csquotes}
\usepackage{mdframed}
\usepackage{enumitem}
\usepackage{ragged2e}

\newmdtheoremenv[%
  outerlinewidth=4pt,
  leftmargin=4em,
  rightmargin=4em,
  innertopmargin=0pt,
  splittopskip=\topskip,
  skipbelow=\baselineskip,
  skipabove=.1\baselineskip]{implication}{Implication}

\begin{document}

\title[ML4SE: A Tertiary Study]{Machine Learning for Software Engineering: A Tertiary Study}

\author{Zoe Kotti}
\author{Rafaila Galanopoulou}
\author{Diomidis Spinellis}
\email{{zoekotti, rgalanopoulou, dds}@aueb.gr}
\orcid{{0000-0003-3816-9162, 0000-0002-5318-9017, 0000-0003-4231-1897}}
\affiliation{%
  \institution{Athens University of Economics and Business}
  \streetaddress{Patission 76}
  \city{Athens}
  \country{Greece}
  \postcode{10434}
}

\renewcommand{\shortauthors}{Z. Kotti \emph{et al.}}

\newcommand{\eg}{\textit{e}.\textit{g}., }
\newcommand{\ie}{\textit{i}.\textit{e}., }
\newcommand{\etal}{\textit{et al}.}
\newcommand{\etc}{\textit{etc}.}
\newcommand{\task}[1]{\textbf{#1}}

\begin{abstract}
Machine learning (ML) techniques increase
the effectiveness of software engineering (SE) lifecycle
activities.
We systematically collected, quality-assessed, summarized, and categorized
83 reviews in ML for SE
published between 2009--2022,
covering 6\,117 primary studies.
The SE areas most tackled with ML are
software quality and testing,
while human-centered areas appear more challenging for ML.
We propose a number of ML for SE research challenges and actions
including:
conducting further empirical validation and industrial studies on ML;
reconsidering deficient SE methods;
documenting and automating data collection and pipeline processes;
reexamining how industrial practitioners distribute their proprietary data;
and implementing incremental ML approaches.
\end{abstract}



\maketitle

\section{Introduction}
\label{sec:intro}
Machine learning (ML) is a thriving discipline
with various practical applications and active research topics,
many of which nowadays entangle the discipline
of software engineering (SE)~\cite{MB18}.
Through ML we can address SE problems
that cannot be completely algorithmically modeled,
or for which existing solutions do not provide satisfactory results yet
(\eg defect/fault detection~\cite{ZWNS06,WAB09,APMP07}).
In addition,
ML finds application in SE tasks where data
cannot be easily analyzed with other algorithms
(\eg software requirements, code comments, code reviews,
issues~\cite{ABDS18,YXLG21,LCB20}).
Another important aspect of ML is that it can significantly reduce manual effort
in common SE tasks
(\eg automatic program repair~\cite{TWBP18}, code suggestion~\cite{GZK18},
defect prediction~\cite{BFEA21}, malware detection~\cite{SB20},
feature location~\cite{CDK15})
with great accuracy results~\cite{WCPM22,SYA22}.
In fields such as health informatics
ML and SE are considered complementary disciplines,
since the growing scale and complexity of healthcare datasets
have posed a challenge for clinical practice and medical research,
requiring new engineering approaches from both fields~\cite{CGDT12}.

In the early nineties,
Huff and Selfridge~\cite{HS90} recognized the need for creating software systems
that partially take some responsibility for their own evolution,
offering the ability to implement, measure, and assess changes easily.
These changes should also contribute
to the overall improvement of the corresponding systems~\cite{Sel93}.
Around the same time,
Brooks~\cite{Bro95} prompted software practitioners
to investigate evolutionary advancements
rather than waiting for revolutionary ones,
since magic solutions are not around the corner.
As a result,
a wave of evolutionary approaches was developed
to address certain inherent essential software challenges including
complexity
(``software entities are more complex for their size
than perhaps any other human construct''),
changeability
(``software is constantly subject to pressures for change''),
conformity
(``software must conform to the human institutions and systems
it comes to interface with''), and
invisibility
(``the reality of software is not inherently embedded in space'')~\cite{Bro87}.

Among the proposed solutions,
ML methods were introduced as a viable alternative
to existing SE approaches,
yielding encouraging results.
Shortly after 2000,
Zhang and Tsai~\cite{ZT03} classified existing applications
of ML methods to SE tasks into seven activity types:
prediction and estimation;
property and model discovery;
transformation;
generation and synthesis;
reuse library construction and maintenance;
requirement acquisition or recovery;
development knowledge management.
Around sixty publications were identified as relevant
and assigned to these categories,
underpinning the increasing trend of employing ML in SE.
However,
as the authors emphasize,
``ML is not a panacea for all the SE problems.
To better use ML methods as tools to solve real-world SE problems,
we need to have a clear understanding of both the problems,
and the tools and methodologies utilized.''
Specifically,
it is vital to be aware of the available ML methods,
their characteristics and theoretical foundations, and
the circumstances under which they are most effectively applied.

Twenty years later,
ML has considerably affected the entire SE lifecycle,
allowing developers to design, develop, and deploy software
in a better, faster, and cheaper manner~\cite{CHMX20}.
In the requirements phase
theorem provers are used
to identify systems with mutually compatible requirements
that can coexist together,
while in the execution phase
multi-objective genetic algorithms can simplify a system's configuration.
ML tools can contribute to a system's monitoring and optimization
by limiting the required cloud resources,
while in testing---the most covered phase by ML
according to a bibliometric analysis by Heradio \etal~\cite{HFCC21}---ML
can automate the prioritization and execution of test suites.
Furthermore,
ML tools can support the automatic identification and repair of software bugs.
By logging these activities,
quality models can be trained to predict useful system features including
software development and issue resolution time,
bug locations, or
software development anti-patterns.

ML for SE, which concerns the application of ML techniques
to SE processes and tools,
should not be confused with the similarly sounding ``SE for ML'' field.
Although the second field's title contains almost the same terms,
its topic is completely different,
namely the application of the SE discipline in the
development and operation of ML applications~\cite{MBFO22}.
Consequently, SE for ML will not concern us further in this review.

To facilitate and assess the impact of ML in SE,
along with the aforementioned study by Zhang and Tsai~\cite{ZT03},
various other secondary reviews have been performed.
Due to the extended associated primary research
that has been published---more than 2\,000 related documents were retrieved
from Elsevier Scopus (see Section~\ref{sec:automated-search}),
secondary reviews usually focus on a particular SE area,
such as software testing (\eg~\cite{DDBE19, CGB20, ZSG16})
or design (\eg~\cite{ZMI17, LSS17}).
Still,
one might wonder what is the overall impact
and current state of the practice of ML in SE,
taking also into account that
some of the ML methods currently employed in SE are rarely encountered
in conventional ML~\cite{MB18}.
To the best of our knowledge,
there is no available tertiary review
systematically summarizing and evaluating
all the published secondary studies in the intersection of the two fields.

This study aims to fill this gap
by methodically collecting, assessing, analyzing, and categorizing
existing secondary research in \textbf{Machine Learning for Software Engineering},
which is commonly abbreviated as \textbf{ML4SE}.
Through the research questions outlined in Section~\ref{sec:objectives-rqs}
we identify
what SE tasks have been tackled with ML techniques,
which SE knowledge areas could be better covered by ML techniques as well as
the prominent ML techniques applied in SE.
We also provide a classification scheme for categorizing ML techniques in SE
along four axes.
Our findings suggest that ML is mainly employed
for automating and optimizing testing;
for predicting software faults, changes, quality, maintainability, and defects;
and
for estimating cost and effort.
Research opportunities lie in the areas of
software construction, configuration management, models and methods.
In addition, we identify
a need for more empirical and industrial studies
evaluating the application of ML techniques in SE.
The majority of secondary reviews
summarize supervised (\ie the training data are labeled),
offline (\ie the system weights are constant),
model-based (\ie patterns are detected in the training data) learning techniques
applied for
classification, clustering, regression,
pattern discovery,
information retrieval, and
generation tasks.

Our research is structured as follows.
In Section~\ref{sec:relwork}
we present related work
comprising all identified tertiary reviews in SE and ML4SE.
Section~\ref{sec:methods} includes a detailed description
of the research objectives, questions, and methods we adopted
to perform this tertiary review.
In Section~\ref{sec:results}
we describe the study findings
with respect to the data extraction process and the research questions,
with their discussion and implications unfolded
in Section~\ref{sec:discussion}.
The study limitations are acknowledged in Section~\ref{sec:threats},
and our final remarks and recommendations for researchers and practitioners
are outlined in Section~\ref{sec:conclusion}.
Following published recommendations~\cite{IHG12},
the code and data\footnote{\url{https://doi.org/10.5281/zenodo.7082429}}
associated with this endeavor are openly available online,
and can be used to perform further empirical studies.

\section{Related work}
\label{sec:relwork}
Tertiary studies (also called \emph{tertiary reviews})
are systematic literature reviews (SLRs)
that aggregate the data and information from a number of existing systematic (secondary) studies
(\eg SLRs, systematic mapping studies, taxonomies) over a specific topic~\cite{KC07}.
There has been a noteworthy increase in the number of tertiary studies for the SE field
since 2007,
when recommended guidelines for performing SLRs in SE were published
by Kitchenham and Charters~\cite{KC07}---in Section~\ref{sec:se}
we present some key efforts.
These help us identify the applicable methods, research questions, and
possible findings.
At the time of writing,
there is only one published tertiary study in ML4SE
(focusing on a specific SE area)---this is summarized in Section~\ref{sec:seml}.

\subsection{Tertiary Studies in SE}
\label{sec:se}
Through a tertiary study
Kitchenham \etal~\cite{KPBP10} describe the status
of SLRs in SE in terms of context and quality,
covering the years 2004--2008, and extending a previous work~\cite{KPBT09}.
The authors report both quantitative and qualitative information
about the identified studies,
such as the corresponding authors' names and institutions, and the addressed SE topics.
The latter are investigated in terms of the knowledge areas (KAs) introduced by the
Guide to the Software Engineering Body of Knowledge (SWEBOK)~\cite{swebok},
and their relation with the courses of the
Curriculum Guidelines for Undergraduate Degree Programs in SE~\cite{Joint04}.
They conclude that there has been an increase in the proportion
of evidence-based SE SLRs.
Most studies tackle general SE topics, and are either
industrial case studies or industrial surveys. With regard to authors,
Magne J{\o}rgensen was the main contributor between 2004--2007, and since then
51 researchers have co-authored up to two reviews each. In terms of origin,
research in Europe has increased,
complementing the previous single presence of US institutions.
The authors argue that,
although the number of published SLRs is increasing,
the majority do not follow an established method.
Nevertheless, the quality of the examined SLRs
abiding by the recommended guidelines has improved.

A number of tertiary reviews examine the evolution of systematic studies
either in the complete SE field or in a specific subfield.
Salma \etal~\cite{SBIN13} highlight that
the Journal of Information and Technology,
the International Conference on Evaluation and Assessment in Software Engineering,
and the Empirical Software Engineering Journal
are the venues with the most significant contributions in SE.
Regarding methodology,
the most troublesome SLR stage appears to be the Search Strategy,
followed by the Data Extraction, and
the Inclusion/Exclusion Criteria.
Da Silva \etal~\cite{SSSC11} report an increase in the number
of SE systematic reviews (particularly of systematic mapping studies)
published between 2004--2009
as well as in the number of covered SE topics.
Still, the quality of reviews seems to remain inferior.
Hanssen \etal~\cite{HSM11} summarized systematic studies
in the area of global software engineering (GSE) investigating agile practices.
Twelve SLRs were identified in GSE between 1990--2009,
with some of them describing agile practices as an evolving trend.
Another study by Marimuthu and Chandrasekaran~\cite{MC17}
presents 60 publications on the topic of software product lines
between 2008--2016,
summarizing their type, quality, authors, publication venue, research topic, and limitations.

In the area of requirements engineering (RE),
Bano \etal~\cite{BZI14} retrieved 53 systematic reviews published between 2006--2014,
and classified them according to the RE subareas.
Non-functional requirements were assessed as the most frequent subarea.
The authors also evaluated the quality of the reviews
using the York University, Centre for Reviews and Dissemination
Database of Abstracts of Reviews of Effects (DARE-4)
criteria,\footnote{\url{https://web.archive.org/web/20070918200401/https://www.york.ac.uk/inst/crd/faq4.htm}}
which we also used in our work---these are presented in Table~\ref{tab:DareCriteria}.
Acknowledged inefficiencies of the reviews concern
unreported or few primary studies,
and inadequately addressed RE subareas.

Distributed software development (DSD---also \emph{global software development}) was another SE area
summarized in the identified tertiary studies.
Alinne \etal~\cite{SJMM12} introduced a systematic tertiary study on
communication in DSD, aiming to
identify and synthesize factors that influence its effectiveness, and
discover its impact on project design.
The authors suggest that more research should be conducted on the topic,
particularly on processes for effectively assessing the maturity of communication
in distributed teams.
In another work,
Marques \etal~\cite{MRC12} collected 14 systematic studies between 2008--2011
discussing the challenges of DSD,
and mapped them with the identified solutions and approaches
that still need further investigation.
Using the SWEBOK KAs,
most studies were categorized in
SE management and process,
software design, and requirements.
With respect to authors and institutions,
it appears that there is considerable cooperation among researchers worldwide.
Finally,
Verner \etal~\cite{VBKT12} enumerated DSD
systematic reviews between 2005--2011, and identified their topics,
active researchers, publication venues, and study quality.

In the area of software testing,
Garousi \etal~\cite{GM16} systematically summarized all state-of-the-art SLRs
published between 1994--2015.
The authors identified
the investigated areas of software testing,
the addressed RQs, and
the citations of the secondary studies, along with
characteristics of the associated primary studies (\eg quality and types).
The tertiary findings reveal a slow improvement in the quality of
the secondary studies over the years.
Regular surveys compose the most frequent type of review,
and also receive significantly more citations than SLRs and
systematic mapping studies.
The most popular testing method seems to be the model-based approach
both in mobile and web services,
while regression and unit testing were assessed
as the most popular testing phases.
There appears to be room for further secondary studies in various testing areas
including test management, beta-testing, exploratory testing,
test stopping criteria, and test-environment development.

\subsection{Tertiary Study in ML4SE}
\label{sec:seml}
In the area of software effort estimation,
Sreekumar \etal~\cite{PMR17} going through 14 SLRs highlight that,
although most studies employ regression-based and ML techniques,
it appears that expert judgment is still preferred by the industry due to its intuitiveness.
The use of ML techniques for effort estimation has been growing since 2017,
combined with analogy-based estimation models.
Concerning accuracy metrics,
there is an increasing use of Mean Magnitude of Relative Error (MMRE),
Median Magnitude of Relative Error (MdMRE), and
Prediction Pred (25\%),
with 78\% of primary studies employing MMRE.
According to the authors,
there is a need for simple comprehensive global models,
due to the distributed nature of software development,
while further research should be conducted
to further improve estimation results derived with ML approaches.

\section{Review Methods}
\label{sec:methods}
\begin{figure}[t]
  \centering
  \includegraphics[width=\textwidth]{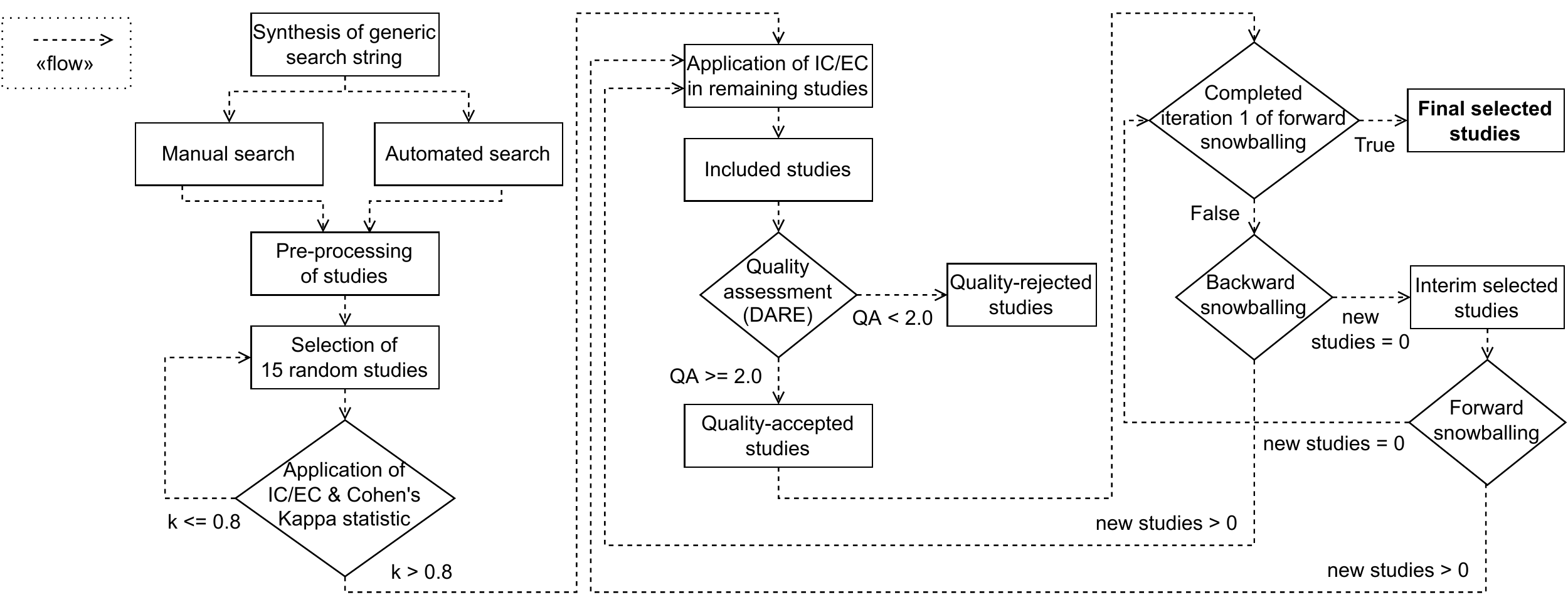}
  \caption{Review Method}
  \Description{Review Method}
  \label{fig:ReviewDiagram}
\end{figure}

To conduct this tertiary review,
we followed the guidelines outlined by Kitchenham and Charters~\cite{KC07}.
A tertiary review employs the same methods
with a typical SLR,
differing in that
primary studies of the latter are considered secondary studies in the former.
Hence,
the review was organized according to the three recommended main phases of an SLR:
\emph{planning},
\emph{conducting}, and
\emph{reporting}.
For the planning phase,
a formal protocol (included in the provided dataset)\footnote{File \emph{review-protocol.md}} was developed and reviewed by all authors,
documenting the review procedures associated with the following processes:
search and selection,
quality assessment,
data extraction,
synthesis, and
analysis.
In all manual activities that required human judgment,
the data extraction and data checking approach was adopted,
as suggested by Brereton \etal~\cite{BKBT07},
where the second author of this paper was the extractor, and
the first was the checker.
The complete review method is presented through a UML information flow diagram
in Fig.~\ref{fig:ReviewDiagram}, after recommended guidelines for systematic studies
to visualize the adopted review process~\cite{VEMA18}.

\subsection{Research Objectives and Questions}
\label{sec:objectives-rqs}
This study aims to:
provide a quality-evaluated catalog of the identified systematic reviews
to the research community;
summarize and assess all published systematic reviews concerning
ML approaches applied in SE activities;
describe the current state of research in ML4SE; and
highlight potential research opportunities in the intersection of the two fields.
To achieve these objectives,
ensuring that the study is comprehensive in its nature,
while providing an in-depth analysis
of the use of ML in SE activities,
the following research questions were defined.

\begin{description}
  \item[RQ1] \emph{What SE tasks have been tackled with ML techniques?}
  \item[RQ2] \emph{What SE knowledge areas could be better covered by ML techniques?}
  \item[RQ3] \emph{What ML techniques have been used in SE?}
\end{description}

By answering these questions we aim to
identify how ML has contributed to SE activities,
uncover SE areas underrepresented by ML, and
identify what ML techniques have been used in the SE activities
described in the collected systematic studies.
Following the practice of other tertiary studies~\cite{KPBP10, BZI14}
we present the quantitative information associated with the examined material
in Section~\ref{sec:data-extraction-res},
rather than dealing with it through a separate research question.
The methods employed to answer each research question
as well as collect and present the related quantitative information
are detailed in Section~\ref{sec:data-extraction}.

\subsection{Search Strategy}
\label{sec:search}
The search strategy was completed in four stages:
automated search in digital sources,
manual search in digital sources,
backward, and forward snowballing.
In the first two stages (depicted at the beginning of Fig.~\ref{fig:ReviewDiagram}),
we searched for studies published between January 2015 and June 2020,
whereas in the last two (visualized at the end of Fig.~\ref{fig:ReviewDiagram}),
earlier and subsequent studies were also examined.

\subsubsection{Automated Search}
\label{sec:automated-search}
\begin{figure}[b]
  \centering
  \includegraphics[scale=0.5]{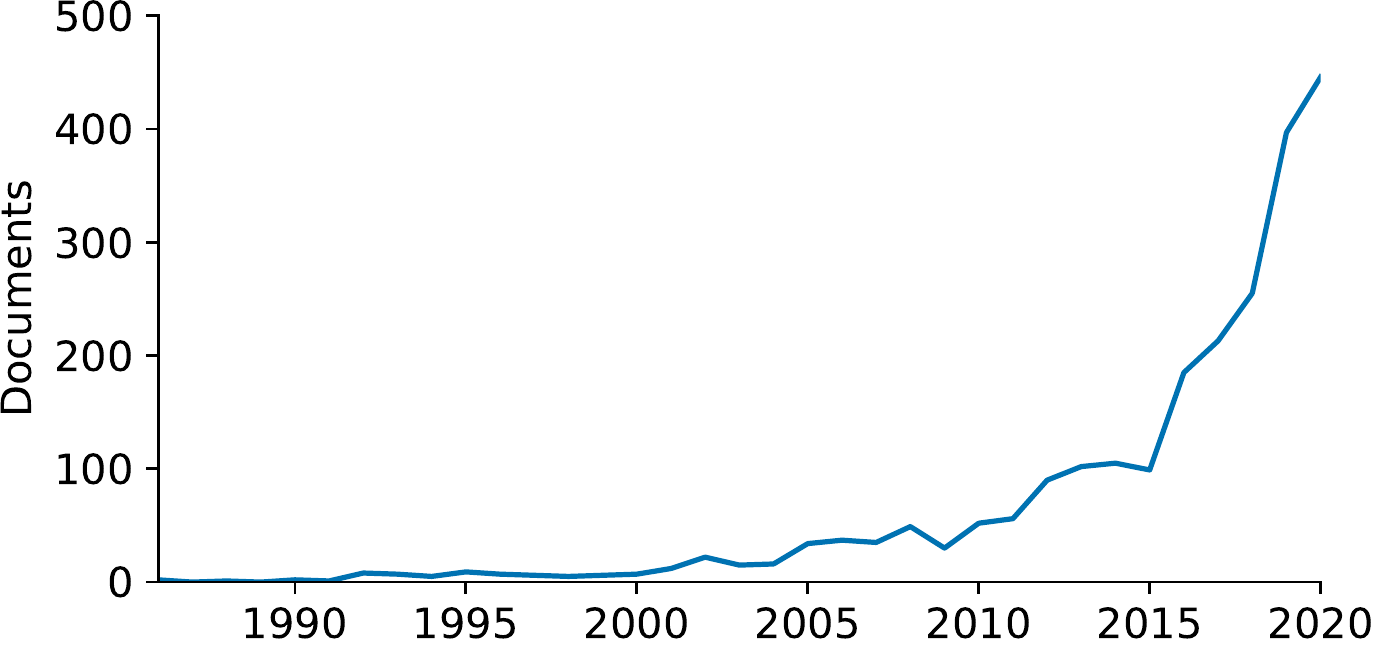}
  \caption{ML and SE documents by year. Adapted from Scopus search results analysis.}
  \label{fig:ScopusAnalysis}
\end{figure}

We selected 2015 as the starting year of the automated search process
for the following reason.
We searched in Elsevier Scopus\footnote{\url{https://www.scopus.com}}
for documents whose title, abstract, or keywords
contained the terms \emph{machine learning} and \emph{software engineering}
up to 2020,
resulting in 2\,316 results.
We then extracted the yearly distribution of these documents in CSV format
from Scopus's \emph{Analyze search results} page,
and visualized them as seen in Fig.~\ref{fig:ScopusAnalysis}.
We observe a constant increase in the number of publications
belonging to the intersection of the two fields after 2015.
Therefore,
we consider this year an inflection point for the joint evolution of the fields,
which would conceivably also mark the appearance of corresponding review surveys.
Studies published outside the selected time window were identified
through repeated snowballing
(see Section~\ref{sec:snowballing}).

The automated search was implemented in two steps.
First, a search string was composed.
Second, this search string was used to systematically query
three online digital libraries:
IEEE Xplore,\footnote{\url{https://ieeexplore.ieee.org/Xplore}}
ACM Digital Library,\footnote{\url{https://dl.acm.org/}} and
Scopus.
We aimed to identify secondary studies in ML4SE,
namely
studies reviewing ML techniques that have been applied in SE activities.
Table~\ref{tab:searchkeys} presents all keywords used in the search string composition,
sorted in three conceptual groups:
keywords related to SE, ML, and secondary studies.
For the SE field,
keywords were derived from the 15 SWEBOK V3 KAs~\cite{swebok}.
The areas of
\emph{Computing Foundations},
\emph{Mathematical Foundations}, and
\emph{Engineering Foundations}
were excluded,
because they were considered outside the study's scope~\cite{UBBM17}.
Despite the V3 guide's relatively old publication year (2014),
all of its KAs remain relevant,
as can be observed from the overview of the ongoing SWEBOK V4~\cite{swebokv4}.
V4 consists of the same KAs,
and redistributes some material to three new KAs:
\emph{Software Architecture},
\emph{Software Operations}, and
\emph{Software Security}~\cite{swebokv4}.

\begin{table}[t]
  \caption{Search Keywords}
  \label{tab:searchkeys}
  \centering
  \resizebox{\textwidth}{!}{
  \begin{tabular}{p{20cm}}
    \textbf{Keywords for SE:}
      Software Configuration Management;
      Software Construction;
      Software Design;
      Software Engineering Economics;
      Software Engineering Management;
      Software Engineering Methods;
      Software Engineering Models;
      Software Engineering Process;
      Software Engineering Professional Practice;
      Software Maintenance;
      Software Quality;
      Software Requirements;
      Software Testing

    \textbf{Keywords for ML:}
      Artificial Neural Networks;
      Bayesian Classifier;
      Classification;
      Clustering;
      Computational Intelligence;
      Computer Learning;
      Data Mining;
      Decision Tree Classification;
      Deep Learning;
      Deep Neural Network;
      Ensemble Learning;
      Generative Adversarial Networks;
      Genetic Programming;
      Grammars;
      Intelligent Computing;
      Learning Algorithms;
      Learning Based;
      Machine Learning-based Applications;
      Machine Learning;
      Meta-Learning;
      Natural Language Processing;
      Regression;
      Reinforcement Learning;
      Semi-supervised Learning;
      Supervised Learning;
      Support Vector Machines;
      Transfer Learning

    \textbf{Keywords for Secondary Studies:}
      analysis of research;
      body of published research;
      centralized tutorial;
      common practices;
      comparative study;
      comprehensive overview;
      conceptual analysis;
      editorial;
      editorial overview;
      editor's preview;
      evidence-based software engineering;
      functional overview;
      general overview;
      in-depth analysis;
      literature analysis;
      literature review;
      literature survey;
      lookup table;
      manifesto;
      meta-analysis;
      meta-survey;
      methodologies;
      overview of existing research;
      past studies;
      review of studies;
      strategic directions;
      structured review;
      study;
      subject matter expert;
      survey and classification;
      survey;
      systematic approach;
      systematic mapping study;
      systematic review;
      taxonomy
  \end{tabular}}
\end{table}

To ensure that the identified studies would be secondary,
we composed a group consisting of 35 keywords adapted from two sources:
a set of 15 keywords introduced in the tertiary study on SLRs in SE
by Kitchenham \etal~\cite{KPBP10};
and a set of 20 keywords manually extracted from the titles of the surveys
published in the ACM Computing Surveys journal.\footnote{\url{https://dl.acm.org/journal/csur}}
Specifically,
we queried the ACM Digital Library
applying the filter \emph{ACM Computing Surveys}
and the ACM CCS 2012~\cite{Rou12} concept
\emph{Surveys And Overviews} (10002944.10011122.10002945).
A set of 304 papers\footnote{File \emph{acm\_comput\_surveys\_overviews.bib}}
were retrieved and exported in BibTeX format.
Next,
titles were isolated and canonicalized
by automatically removing all stop words and punctuation.
The second author examined the canonicalized titles by hand
to identify any additional terminology used for describing secondary studies
that was not already part of the 15 aforementioned keywords.
As a result,
20 additional keywords were included in the group of keywords for secondary studies.

Equivalently,
the third group consists of 27 keywords manually extracted
from the titles of the ML-related surveys of the ACM Computing Surveys journal.
Again,
we queried the ACM Digital Library
applying the filter \emph{ACM Computing Surveys}
and the ACM CCS 2012 concept \emph{Machine Learning} (10010147.10010257).
In total,
148 papers\footnote{File \emph{acm\_comput\_ml\_surveys.bib}}
were fetched and
their titles were canonicalized.
Similarly,
the second author inspected the canonicalized titles by hand
to identify the ML terminology used in secondary studies,
resulting in the 27 keywords listed in Table~\ref{tab:searchkeys}.

All possible 3-tuples occurring from these three groups were used
to compose search strings and
query the fields of document title, abstract, and author keywords in the digital libraries.\footnote{File \emph{dl\_search\_queries.txt}}
Since each library has its own syntax,
the procedure was adjusted accordingly.
In the end,
a set of 1\,897 studies were collected,
from which duplicates were removed based on the studies' digital object identifiers
keeping the latest occurrences,
resulting in 1\,566 unique studies.\footnote{File \emph{dl\_search\_results.csv}}

\subsubsection{Manual Search}
\label{sec:manual-search}
To increase the coverage of the automated search process,
we additionally performed a manual search in each digital library
using one random search string of all possible 3-tuples.
In this way, one more relevant paper was found (1\,567 studies in total).

\subsubsection{Backward and Forward Snowballing}
\label{sec:snowballing}
After completing the quality assessment of the secondary reviews
described below in Section~\ref{sec:qa},
the backward snowballing procedure~\cite{Woh14} was applied
on their referenced studies.
Following the data extraction and data checking process,
a set of 3\,195 studies referenced by the quality-accepted reviews were evaluated
using the inclusion and exclusion criteria (IC/EC) outlined in Section~\ref{sec:ic_ec},
after reading their title, keywords, and abstract.\footnote{File \emph{backward\_snowballing\_references.csv}}
In this way,
16 additional secondary studies were included and quality-assessed,
with the earliest study having been published in 2003.\footnote{File \emph{backward\_snowballing.csv}}
Out of these 16 studies,
seven passed the quality assessment and made it into the interim set of reviews
(Fig.~\ref{fig:ReviewDiagram}).
One more backward snowballing round was performed on these seven studies,
which did not result in any further relevant reviews.

A single iteration of forward snowballing~\cite{WMRK20} was performed
on the quality-accepted studies that occurred from the initial search and
the backward snowballing process.\footnote{One iteration is sufficient according to the SLR guidelines by Wohlin \etal~\cite{WMRK20}.}
Although Wohlin \etal~\cite{WMRK20} recommend Google Scholar as a search engine
(compared to IEEE Xplore and ACM Digital Library),
we opted for Scopus which has a satisfactory citation coverage~\cite{MOTL18},
provides automated search functionality, and
supports citation information extraction in CSV format.
In contrast,
extracting citations from Google Scholar by hand
would entail excessive manual effort,
hinder reproducibility, and
might involve missed records from the human raters~\cite{PVK15}.
Consequently,
we automatically retrieved from Scopus on June 29th, 2022 a total of 2\,461 studies
referencing the quality-accepted reviews, removed duplicates, and evaluated them
following the same process we used
for backward snowballing.\footnote{Files \emph{forward\_snowballing\_reviewer\_\{1,2\}.csv}}
An additional set of 84 studies were included, from which 43 were quality-accepted.
Two of these accepted studies extended accepted studies from the initial search,
and we kept the extensions as they are more complete.\footnote{The review by Gon\c{c}ales \etal~\cite{GFS21} extends~\cite{GFSF19}, and the review by Idri \etal~\cite{IHA16} extends~\cite{IHA16b}.}

\subsection{Selection Criteria}
\label{sec:ic_ec}
The following set of IC/EC were applied to all studies collected
with the search strategy (Section ~\ref{sec:search})
to ensure that only relevant secondary studies would be included in the tertiary review.

\par{\emph{Inclusion Criteria.}}
\begin{itemize}
  \item Only secondary studies
  (\ie SLRs, systematic mapping studies, meta-analyses)
  conducted with documented systematic methods
  (defined research questions, search process,
  data extraction and presentation) are included.
  \item Taxonomies with the following planning characteristics~\cite{UBBM17} are included.
    I) A particular SWEBOK KA is examined.
    II) The objectives and scope of research are clearly defined.
    III) The subject matter of classification
    (units of classification, classes, categories) is described.
    IV) A specific classification structure type
    (hierarchy, tree, paradigm, facet-based) is selected.
    V) A classification procedure type
    (qualitative, quantitative, or both) is defined.
    VI) The data sources and collection methods are documented.
  \item Publications reporting results on the use of
  ML techniques in SE activities are included.
\end{itemize}

\par{\emph{Exclusion Criteria.}}
\begin{itemize}
  \item Non-secondary studies are excluded. These include:
  empirical studies;
  experimental evaluation studies;
  comparative studies with experimental results;
  reports and summaries of workshops
  (\eg~\cite{TVMN20, LZLL15, AENK16, CH19, FPAA20});
  non-implemented future research plans
  (\eg~\cite{Yin15}).
  \item Publications mentioning the use of ML
  without describing the employed techniques are excluded.
  \item Inaccessible papers and reports (\ie only the abstract is available) are excluded.
  \item Publications that are not written in English are excluded
  (\eg~\cite{OT16, NGBL19}).
  \item Informal literature surveys (\ie with undefined research questions,
  undocumented search, data extraction or analysis processes) are excluded.
\end{itemize}

\subsection{Selection Process}
\label{sec:selection-process}
After collecting the candidate secondary studies
through the search strategy described in Section~\ref{sec:search},
we manually applied the IC/EC (Section~\ref{sec:ic_ec})
to exclude irrelevant instances.
We also included three more studies
\cite{ABDS18,YXLG21,LCB20}
that, though not identified through the search strategy,
were suggested during the paper's review and
satisfied the other selection criteria.

Aligning with the adopted guidelines~\cite{KC07},
the selection process was based on the papers' title, author keywords, and abstract,
and was a two-step process (depicted in the middle of Fig.~\ref{fig:ReviewDiagram}).
First, the data extractor and data checker reviewed a set of 15 randomly selected studies,
and determined their inclusion or exclusion.\footnote{File \emph{cohen\_kappa\_agreement.csv}}
Their level of agreement (inter-rater reliability) on this set was measured
using Cohen's Kappa statistic~\cite{PDGT20}, and
any discrepancies that occurred were resolved by consensus~\cite{KC07, KPBP10, MBFO22}.
This step was repeated until a score of at least 0.8 was reached---in our case,
only one iteration was needed,
in which 14 studies were excluded and one was included by both authors.
In this way,
we ascertained that both researchers agreed on the IC/EC,
allowing them to individually review the abundant remaining studies
of the automated and manual search
($n=$ 1\,552---Sections~\ref{sec:automated-search},~\ref{sec:manual-search})
with high inter-rater reliability.

Consequently,
in the second step,
the remaining papers were split in two,
and each of the first two authors individually reviewed a half ($n=$ 776),
again based on the IC/EC.\footnote{Files \emph{study\_selection\_reviewer\_\{1,2\}.csv}}
In case the authors could not determine a paper's inclusion
only by its title, keywords, or abstract,
the full text was consulted.
The inclusion of a limited number of studies,
whose scope remained miscellaneous even after the full-text reading,
was determined after discussion between the first two authors.
The same method was adopted for the backward and forward snowballing processes
(Section~\ref{sec:snowballing}).
Eventually,
from all searches,
a set of 140 distinct secondary studies were selected and quality-assessed,
as detailed in the following Section~\ref{sec:qa}.

\subsection{Quality Assessment}
\label{sec:qa}

We manually assessed the quality of the 140 selected secondary reviews
to ensure concreteness of our study results.
For this,
we followed a recommended quality assessment process for tertiary studies~\cite{KPBT09, KB13}
using the DARE-4 criteria introduced in Section~\ref{sec:relwork},
and presented in Table~\ref{tab:DareCriteria}.
These are recommended criteria for assessing the quality of tertiary studies
by Kitchenham \etal~\cite{KPBP10},
and are also the most commonly used ones in SE tertiary studies~\cite{CFFQ21}.
Although there is a more recent version of
the DARE criteria (\ie DARE-5),\footnote{\url{https://www.crd.york.ac.uk/CRDWeb/AboutPage.asp} (\emph{About DARE})}
we opted for DARE-4,
which is mostly similar to DARE-5 apart from Criterion 3
(\emph{Were the included studies synthesized?}),
which is not clearly prescribed and we did not consider relevant to the goals of our study.

\begin{table}[t]
  \caption{DARE-4 Criteria for Quality Assessment}
  \label{tab:DareCriteria}
  \centering
  \resizebox{\textwidth}{!}{
  \begin{tabular}{l l r l}
    \hline
    \textbf{QA Criterion} & \textbf{Assessment} & \textbf{Score} & \textbf{Description} \\
    \hline
    \multirow{3}{*}{1. IC/EC} & Yes & 1 & Explicit definition of IC/EC \\
      & Partial & 0.5 & Implicit definition of IC/EC \\
      & No & 0 & No IC/EC defined \\
      \hline
    \multirow{3}{*}{2. Search space} & Yes & 1 & 4+ digital libraries searched and additional search strategies applied \\
      & Partial & 0.5 & 3--4 digital libraries searched, but no extra search strategies applied \\
      & No & 0 & 1--2 digital libraries searched or very restricted search \\
    \hline
    \multirow{3}{*}{3. Quality assessment of primary studies} & Yes & 1 & Quality criteria explicitly described and applied \\
      & Partial & 0.5 & Implicit quality assessment \\
      & No & 0 & No quality assessment \\
    \hline
    \multirow{3}{*}{4. Information regarding primary studies} & Yes & 1 & Complete information presented about primary studies \\
      & Partial & 0.5 & Summary information presented about primary studies \\
      & No & 0 & Results of primary studies not specified \\
    \hline
  \end{tabular}}
\end{table}

The DARE-4 criteria are based on four questions.
Kitchenham \etal~\cite{KPBP10} refined them
by attributing points to each criterion.
In this way
each criterion is scored either as Y (\emph{yes}---1 point),
P (\emph{partially}---0.5 point),
or N (\emph{no}---0 points).
The total score assigned to a study is the aggregate of the scores of all four questions.
Thus,
the highest possible score for a study is four, while the lowest is zero.
Included studies should receive a score of at least two.

Again, we followed the data extraction and data checking approach,
reaching an inter-rater agreement of 82\%.\footnote{File \emph{dare\_assessment.csv}}
The majority of disagreements occurred in the last question (QA4),
which concerns the provided information about the reviewed primary studies---we
attribute this to the higher entailed subjectivity of the particular question.
Through this process, 57 out of 140 (41\%) studies were excluded,
with a total score less than two.
The total scores of accepted studies are presented
in Tables~\ref{tab:dataextractiontab},~\ref{tab:dataextractiontab2}.
We observed that
excluded secondary studies with inferior quality do not explicitly
document their IC/EC (QA1) or search sources (QA2),
or fail to assess the quality of the included primary studies (QA3).

\subsection{Data Extraction}
\label{sec:data-extraction}
The information extracted from each quality-accepted secondary study was the following.
\begin{itemize}[leftmargin=*]
  \item Title and source
  (journal, workshop proceedings, conference proceedings, book chapter)
  \item Publication year---to outline the annual evolution and research interest
  in ML4SE.
  \item Publication venue---to highlight prominent publishers in the particular area.
  \item Author names, institutions, and countries---to discover leading research teams.
  \item Study type (\eg SLR, systematic mapping study, taxonomy)
  \item Research method---to investigate guidelines highly adopted by secondary
  studies.
  \item Quality assessment score
  \item Number of primary studies---to approximate how many
  primary studies are implicitly covered by our tertiary study.
  \item Application domain in terms of SWEBOK KAs and subareas
  as well as SE tasks covered by each secondary study---to answer RQ1 and RQ2.
  \item Implications for further research and comments
  concerning the use of ML in SE---to answer RQ2.
  \item Employed ML techniques---to answer RQ3.
\end{itemize}

To extract the aforementioned information
that was needed to answer the RQs introduced in Section~\ref{sec:objectives-rqs},
we employed the following methods.

\par{\emph{RQ1: What SE tasks have been tackled with ML techniques?}}
To answer this,
we extracted from each secondary study its application domain in terms of
related SWEBOK KA, subarea, and SE task(s).\footnote{File \emph{knowledge\_areas.csv}}
When a study was associated with multiple KAs/subareas,
the most prominent one was kept
(\ie the most covered KA/subarea by the included primary studies,
or the most analyzed one by the review authors).
The process was implemented following the data extraction and data checking approach,
and any conflicts that occurred were resolved by the last author.
For the extraction of the SE tasks,
we followed the open coding practice~\cite{CS90}
by manually applying codes
(\ie SE tasks---\eg \emph{test automation},
\emph{software maintainability prediction},
\emph{software defect prediction},
\emph{bug prioritization}) to the studies.
The study list was split in two,
and each of the first two authors individually applied codes (SE tasks) to a half
(in a shared online spreadsheet).
To maintain consistency between the two authors,
the codes were mainly extracted from the study's title, author keywords, abstract,
or introduction.
Next, the authors discussed and grouped together conceptually-related codes
by generalizing or specializing them,
employing the Qualitative Content Analysis approach~\cite{LSMH15}.
In the end,
each secondary study was associated with at least one and up to three SE tasks.
Consequently,
\textbf{a SE task may be associated with multiple KAs}.

\par{\emph{RQ2: What SE knowledge areas could be better covered by ML techniques?}}
For this,
we used the results of RQ1 to identify SWEBOK KAs
that are insufficiently covered by ML techniques.
Moreover, we extracted by hand any implications for further research
as well as comments regarding the use of ML in SE
that were mentioned in the associated reviews.\footnote{File \emph{further\_research.csv}}
To do this,
we searched the sections of abstract, introduction, results, conclusion,
and further research or future directions (where available)
of all secondary studies to identify ML-related research opportunities in each KA.
Lastly, we extracted from the same sections
any identified issues or obstacles associated with the use of ML techniques in SE.

\par{\emph{RQ3: What ML techniques have been used in SE?}}
To address this,
we classified each secondary study using a classification scheme,
again following the data extraction and data checking approach.\footnote{File \emph{ml\_techniques.csv}}
The classification scheme was constructed from two sources
and consists of four axes:
the role of AI in SE~\cite{Har12},
the supervision type~\cite{KJ18},
the incrementality type~\cite{KJ18}, and
the generalizability type~\cite{KJ18}.
The \emph{role of AI in SE} includes the following three categories.

\begin{itemize}[leftmargin=*]
  \item Computational search and optimisation techniques
  (the field known as Search Based Software Engineering---SBSE):
  The goal of this area is to reconstruct SE problems
  as optimization problems,
  which can then be tackled with computational search-related AI techniques.
  \item Fuzzy and probabilistic methods for reasoning
  in the presence of uncertainty:
  AI techniques are used to address real-world problems,
  which are inherently fuzzy and probabilistic
  (\eg the use of Bayesian probabilistic reasoning
  to model software reliability or analyze users).
  \item Classification, learning and prediction:
  This area involves the application of ML techniques
  such as artificial neural networks,
  case-based reasoning,
  and rule induction
  to model and predict SE tasks
  (\eg software project prediction, ontology learning, defect prediction).
\end{itemize}

\emph{Supervision} expands to
supervised,
unsupervised,
semi-supervised, and
reinforcement learning.
In supervised learning,
the training dataset used by the ML-based system includes the desired solutions
(\ie labels),
whereas in unsupervised learning,
the training dataset is unlabeled.
Semi-supervised learning involves many unlabeled data
combined with a few labeled instances,
while reinforcement learning is concerned
with learning to control a system,
in order to maximize a numerical performance measure
that expresses a long-term objective~\cite{Sze10}.
Contrary to supervised,
in reinforcement learning,
only partial feedback is provided to the system
about its predictions.

The third axis, \emph{incrementality}, consists of
online/incremental and batch/offline learning.
In online learning,
the ML algorithms are trained incrementally
by feeding them data instances sequentially on the fly (as they arrive),
either individually or in small groups (\ie mini-batches).
These algorithms constantly update the system weights;
thus the error calculation uses different weights for each input sample.
On the other hand, in offline learning,
the ML algorithms are first trained with all available training data,
and are then released into production
without the ability to learn incrementally---they
only apply what they have learned.
These algorithms keep the system weights constant
while computing the error associated with each input sample.

The axis of \emph{generalizability} expands to
instance-based and model-based learning.
In instance-based learning,
the system stores the data and generalizes to new instances
by employing a similarity measure,
whereas in model-based learning,
patterns are detected in the training data,
and are used to build a predictive model.
In all axes,
studies were classified to the most prominent category
(when more than one categories could be mapped).
In addition,
we extracted by hand the ML techniques
employed in the primary studies,
when reported in the corresponding secondary,
to determine the most popular ones for SE tasks.

\section{Results}
\label{sec:results}
In this section we present our research findings
in regard to the data extraction process described in Section~\ref{sec:data-extraction}
and the research questions outlined in Section~\ref{sec:objectives-rqs}.

\subsection{Data Extraction}
\label{sec:data-extraction-res}
The papers in our final set of 83 quality-accepted secondary studies
were published between 2009--2022,
and cover 6\,117 non-unique primary studies
(conference papers, journal papers, theses, and technical reports)
published between 1990--2021.
The majority of secondary reviews ($n=$ 63; 76\%) were published in journals
as opposed to conference proceedings ($n=$ 20; 24\%).
An overview of all reviews sorted by publication year is presented
in Tables~\ref{tab:dataextractiontab},~\ref{tab:dataextractiontab2}.

\begin{table}[t]
  \caption{Overview of Secondary Studies (1/2)}
  \label{tab:dataextractiontab}
  \centering
  \resizebox{\textwidth}{!}{
  \begin{tabular}{rlrlrrr}
    \hline
    \textbf{Study}   & \textbf{Venue}                     & \textbf{Year}  & \textbf{Publisher}                                 & \textbf{QA Score}  & \textbf{Primary}  & \textbf{Covered Years} \\
    \hline
    \cite{RMT09}     & ESEM '09                           & 2009           & IEEE                                               & 4.0                & 15                & 1994--2008             \\
    \cite{Rad10}     & Int. J. Soft. Eng. Comput.         & 2010           & International Science Press                        & 2.5                & 23                & 2002--2010             \\
    \cite{HBBG12}    & IEEE Trans. Softw. Eng.            & 2012           & IEEE                                               & 4.0                & 36                & 2002--2010             \\
    \cite{WLLH12}    & Inf. Softw. Technol.               & 2012           & Elsevier                                           & 4.0                & 84                & 1992--2010             \\
    \cite{BRA14}     & Empir. Softw. Eng.                 & 2014           & Springer                                           & 3.0                & 79                & 1999--2011             \\
    \cite{BKS15}     & J. Syst. Softw.                    & 2015           & Elsevier                                           & 3.0                & 13                & 2005--2014             \\
    \cite{AS15}      & Requir. Eng.                       & 2015           & Springer                                           & 4.0                & 29                & 1999--2013             \\
    \cite{IAA15}     & SNPD '15                           & 2015           & IEEE                                               & 3.5                & 35                & 2000--2013             \\
    \cite{Mal15}     & Appl. Soft Comput.                 & 2015           & Elsevier                                           & 4.0                & 64                & 1995--2013             \\
    \cite{MB15}      & ICRITO '15                         & 2015           & IEEE                                               & 2.0                & 21                & 1998--2014             \\
    \cite{CTH16}     & Empir. Softw. Eng.                 & 2016           & Springer                                           & 2.0                & 167               & 1999--2014             \\
    \cite{MC16}      & Int. J. Softw. Eng. Knowl. Eng.    & 2016           & World Scientific Publishing                        & 4.0                & 96                & 1991--2015             \\
    \cite{IHA16}     & J. Syst. Softw.                    & 2016           & Elsevier                                           & 3.0                & 24                & 2000--2016             \\
    \cite{WSCW16}    & ICSME '16                          & 2016           & IEEE                                               & 3.5                & 29                & 2000--2015             \\
    \cite{KIA16}     & SEAA '16                           & 2016           & IEEE                                               & 3.5                & 19                & 1997--2015             \\
    \cite{MB16}      & Int. J. Comput. Appl. Technol.     & 2016           & Inderscience Publishers                            & 2.0                & 21                & 1998--2011             \\
    \cite{SLLD16}    & SNPD '16                           & 2016           & IEEE                                               & 2.0                & 38                & 2007--2015             \\
    \cite{ZSG16}     & J. Syst. Softw.                    & 2016           & Elsevier                                           & 3.0                & 79                & 2005--2015             \\
    \cite{AZAB17}    & IEEE Access                        & 2017           & IEEE                                               & 3.5                & 103               & 2005--2016             \\
    \cite{DLLW17}    & APSEC '17                          & 2017           & IEEE                                               & 2.5                & 40                & 1998--2016             \\
    \cite{KJMG17}    & ACM Comput. Surv.                  & 2017           & ACM                                                & 2.5                & 47                & 2007--2015             \\
    \cite{MKR17}     & Swarm Evol. Comput.                & 2017           & Elsevier                                           & 4.0                & 78                & 1992--2015             \\
    \cite{LSS17}     & SPLC '17                           & 2017           & ACM                                                & 3.5                & 25                & 2005--2017             \\
    \cite{TBA17}     & Softw. Qual. J.                    & 2017           & Springer                                           & 3.5                & 10                & 2012--2014             \\
    \cite{UGDN17}    & Artif. Intell. Rev.                & 2017           & Springer                                           & 2.5                & 32                & 2003--2015             \\
    \cite{ZMI17}     & ICET '17                           & 2017           & IEEE                                               & 2.0                & 22                & 1998--2016             \\
    \cite{AY18}      & CTCEEC '17                         & 2018           & IEEE                                               & 2.5                & 17                & 2006--2014             \\
    \cite{OT18}      & J. Syst. Softw.                    & 2018           & Elsevier                                           & 4.0                & 52                & 2000--2016             \\
    \cite{ABDS18}    & ACM Comput. Surv.                  & 2018           & ACM                                                & 2.0                & 91                & 2007--2018             \\
    \cite{EDR18}     & SPICE '18                          & 2018           & Springer                                           & 2.5                & 25                & 1998--2017             \\
    \cite{OQCH19}    & CITT '18                           & 2018           & Springer                                           & 2.0                & 20                & 2013--2016             \\
    \cite{RLJA18}    & IEEE Access                        & 2018           & IEEE                                               & 4.0                & 113               & 1993--2016             \\
    \cite{SS18}      & J. Syst. Softw.                    & 2018           & Elsevier                                           & 3.0                & 445               & 1996--2016             \\
    \cite{MDDC19}    & Comput. Electr. Eng.               & 2019           & Elsevier                                           & 4.0                & 31                & 2002--2017             \\
    \cite{AG19b}     & SEAA '19                           & 2019           & IEEE                                               & 3.5                & 30                & 2007--2018             \\
    \cite{CRCP19}    & Int. J. Softw. Eng. Knowl. Eng.    & 2019           & World Scientific Publishing                        & 3.0                & 26                & 1999--2016             \\
    \cite{APSW19}    & Inf. Softw. Technol.               & 2019           & Elsevier                                           & 3.5                & 15                & 2000--2017             \\
    \cite{AR19}      & ICECTA '19                         & 2019           & IEEE                                               & 3.0                & 15                & 2007--2017             \\
    \cite{DDBE19}    & IEEE Trans. Reliab.                & 2019           & IEEE                                               & 3.0                & 48                & 1995--2018             \\
    \cite{HTG19}     & IEEE Trans. Softw. Eng.            & 2019           & IEEE                                               & 4.0                & 30                & 2008--2015             \\
    \cite{SPKK19}    & Symmetry                           & 2019           & MDPI                                               & 3.0                & 98                & 1995--2018             \\
    \cite{ECIA19}    & e-Inform. Softw. Eng. J.           & 2019           & Wroc{\l{}}aw University of Science and Technology  & 3.5                & 82                & 2000--2018             \\
    \cite{MK19}      & e-Inform. Softw. Eng. J.           & 2019           & Wroc{\l{}}aw University of Science and Technology  & 3.0                & 38                & 2000--2019             \\
    \cite{AG19a}     & J. Softw.: Evol. Process           & 2019           & Wiley                                              & 4.0                & 75                & 1991--2017             \\
    \cite{KBAA19}    & IEEE Access                        & 2019           & IEEE                                               & 3.0                & 58                & 2016--2019             \\
    \cite{NZM19}     & ICCSRE '19                         & 2019           & IEEE                                               & 2.5                & 46                & 1995--2017             \\
    \cite{TAB19}     & ICETC '19                          & 2019           & ACM                                                & 2.0                & 31                & 2003--2019             \\
    \cite{AHS20}     & ICPC '20                           & 2020           & ACM                                                & 2.0                & 33                & 2012--2019             \\
    \cite{CA20}      & SEAA '20                           & 2020           & IEEE                                               & 2.5                & 38                & 2009--2019             \\
    \cite{CS20}      & SEAA '20                           & 2020           & IEEE                                               & 2.5                & 196               & 2012--2017             \\
    \cite{AG20}      & ICOSST '20                         & 2020           & IEEE                                               & 3.5                & 34                & 2007--2019             \\
    \cite{SB20}      & IEEE Access                        & 2020           & IEEE                                               & 3.0                & 32                & 2009--2019             \\
    \cite{SZWA20}    & IET Softw.                         & 2020           & IET                                                & 3.5                & 28                & 2014--2020             \\
    \cite{AFKK20}    & Secur. Commun. Netw.               & 2020           & Wiley                                              & 2.5                & 12                & 2011--2019             \\
    \cite{LCB20}     & ACM Comput. Surv.                  & 2020           & ACM                                                & 2.0                & 267               & 1992--2019             \\
    \hline
  \end{tabular}}
\end{table}

\begin{table}[t]
  \caption{Overview of Secondary Studies (2/2)}
  \label{tab:dataextractiontab2}
  \centering
  \resizebox{\textwidth}{!}{
  \begin{tabular}{rlrlrrr}
    \hline
    \textbf{Study}   & \textbf{Venue}                     & \textbf{Year}  & \textbf{Publisher}                                 & \textbf{QA Score}  & \textbf{Primary}  & \textbf{Covered Years} \\
    \hline
    \cite{CGB20}     & SAC '20                            & 2020           & ACM                                                & 2.0                & 320               & 2017--2019             \\
    \cite{ML20}      & Soft Comput.                       & 2020           & Springer                                           & 3.5                & 36                & 1993--2019             \\
    \cite{PCCR20}    & Inf. Softw. Technol.               & 2020           & Elsevier                                           & 3.0                & 93                & 2004--2015             \\
    \cite{LLN20}     & ACM Comput. Surv.                  & 2020           & ACM                                                & 2.0                & 109               & 1999--2019             \\
    \cite{AAA20}     & Arab. J. Sci. Eng.                 & 2020           & Springer                                           & 4.0                & 17                & 2005--2018             \\
    \cite{ECIA20}    & J. Comput. Sci. Technol.           & 2020           & Springer                                           & 3.5                & 77                & 2000--2018             \\
    \cite{PAMJ21}    & J. Syst. Softw.                    & 2021           & Elsevier                                           & 2.5                & 69                & 2005--2019             \\
    \cite{YXLG21}    & ACM Comput. Surv.                  & 2021           & ACM                                                & 2.0                & 250               & 2006--2020             \\
    \cite{BFEA21}    & J. Comput. Sci.                    & 2021           & Science Publications                               & 3.0                & 40                & 2016--2020             \\
    \cite{MAAK21}    & Intell. Autom. Soft Comput.        & 2021           & Tech Science Press                                 & 2.5                & 22                & 2016--2019             \\
    \cite{RAJH21}    & Int. J. Adv. Comput. Sci. Appl.    & 2021           & The Science and Information Organization           & 2.5                & 48                & 2017--2020             \\
    \cite{SUBG21}    & ACM Comput. Surv.                  & 2021           & ACM                                                & 2.5                & 92                & 2010--2019             \\
    \cite{AAG21}     & J. Softw.: Evol. Process           & 2021           & Wiley                                              & 3.5                & 145               & 1993--2018             \\
    \cite{AHC21}     & J. Softw.: Evol. Process           & 2021           & Wiley                                              & 4.0                & 31                & 2011--2019             \\
    \cite{SGG21}     & Empir. Softw. Eng.                 & 2021           & Springer                                           & 2.0                & 111               & 2009--2020             \\
    \cite{ZZA21}     & REW '21                            & 2021           & IEEE                                               & 3.5                & 65                & 2010--2020             \\
    \cite{SSB21}     & SN Comput. Sci.                    & 2021           & Springer                                           & 3.0                & 30                & 1995--2020             \\
    \cite{KIJS21}    & IEEE Access                        & 2021           & IEEE                                               & 3.0                & 110               & 2004--2021             \\
    \cite{PMT21}     & Expert Syst. Appl.                 & 2021           & Elsevier                                           & 4.0                & 154               & 1990--2019             \\
    \cite{ANA21}     & Sci. Comput. Program.              & 2021           & Elsevier                                           & 4.0                & 75                & 1993--2019             \\
    \cite{GFS21}     & Inf. Softw. Technol.               & 2021           & Elsevier                                           & 2.5                & 63                & 2010--2020             \\
    \cite{MKAK22}    & Softw.: Pract. Exp.                & 2022           & Wiley                                              & 4.0                & 35                & 1997--2020             \\
    \cite{WCPM22}    & ACM Trans. Softw. Eng. Methodol.   & 2022           & ACM                                                & 3.0                & 128               & 2009--2019             \\
    \cite{YXLB22}    & ACM Trans. Softw. Eng. Methodol.   & 2022           & ACM                                                & 3.0                & 421               & 2009--2020             \\
    \cite{BK22}      & Comput. Electr. Eng.               & 2022           & Elsevier                                           & 4.0                & 68                & 2010--2021             \\
    \cite{SYA22}     & IEEE Access                        & 2022           & IEEE                                               & 2.0                & 62                & 2016--2021             \\
    \cite{PAKS22}    & Eng. Appl. Artif. Intell.          & 2022           & Elsevier                                           & 2.0                & 146               & 2009--2020             \\
    \cite{LM22}      & Stud. Syst. Decis. Control         & 2022           & Springer                                           & 2.5                & 45                & 2005--2020             \\
    \hline
  \end{tabular}}
\end{table}

\par{\textbf{Top authors}}
In total,
274 researchers contributed to the 83 secondary studies.
Between 2009--2022,
the most active researcher in the field was Ruchika Malhotra,
having co-authored six studies,
followed by Alain Abran and Ali Idri.

\par{\textbf{{Top institutions}}
The studies originate from 140 institutions.
Delhi Technological University (Delhi, India) is on the top of the list with seven studies,
followed by \'{E}cole de Technologie Sup\'{e}rieure (Montreal, Canada),
Mohammed V University (Rabat, Morocco), and
University of Adelaide (Adelaide, Australia).

\par{\textbf{Distribution of studies}}
Figure~\ref{fig:PubYear} depicts the number of publications by year and publisher.
Most secondary studies were published between 2019--2021,
while no studies were found from 2011 and 2013.
Overall,
we observe a notable increase in the number of studies after 2015,
which aligns with the Scopus results visualized in Fig.~\ref{fig:ScopusAnalysis}.
Regarding publishers' distribution, IEEE is first with 25 publications,
followed by Elsevier with 17, Springer with 13,
ACM with twelve, and
Wiley with five studies.

\par{\textbf{Quality of studies}}
As deduced from Fig.~\ref{fig:QualityYear},
there seems to be a fixed average quality of secondary studies after 2014.
The total average score remains above average (2) each year,
implying an overall adequate (yet not perfect) quality of secondary studies.
Although few studies were published in 2009 and 2012,
these had the highest scores in all questions,
suggesting that the DARE-4 criteria (Table~\ref{tab:DareCriteria})
have been systematically adopted from early on.

\par{\textbf{Research types of studies}}
The majority of secondary studies ($n=$ 53; 64\%) are primarily SLRs,
while 19\% ($n=$ 16) are systematic mapping studies,
16\% ($n=$ 13) are surveys, and
a single study is a taxonomy.
Seven studies include a second research type:
SLR (together with systematic mapping study)~\cite{OT18, ECIA19}, and
meta-analysis (with SLR as primary)~\cite{HTG19,ECIA20,MDDC19,APSW19,SB20}.

\par{\textbf{Research methods of studies}}
The most commonly adopted guidelines for SLRs and surveys are those
by Kitchenham \etal~\cite{KC07, Kit04, KBB15},
while most systematic mapping studies follow the guidelines by Petersen \etal~\cite{PFMM08, PVK15},
and Kitchenham \etal~\cite{KBB11}.
In addition,
some studies employ the structure proposed by Hall \etal~\cite{HBBG12}
for conducting reviews and presenting results.
The snowballing search method by Wohlin \etal~\cite{WRHO12, WP13, Woh14} is also used
in some studies complementary to the aforementioned guidelines,
based on published recommendations
regarding the inclusion of manual target searches in systematic reviews~\cite{JS07}.
To compose the research questions,
some reviews adopt recommendations by Easterbrook \etal~\cite{ESSD08}, and
Sabir \etal~\cite{SPRG19}.
To assess the quality of primary studies,
various criteria have been used, such as
the ones by Zhou \etal~\cite{ZZH15} and Dyb\r{a} \etal~\cite{DD08}, and
the Systematic Review Checklist
by the Critical Appraisal Skills Programme (CASP)~\cite{Cas19}.
Moreover,
a variety of methods are used for data synthesis, analysis, and visualization, including
content analysis~\cite{Kri18},
grounded theory~\cite{Cha14}, and
box plots~\cite{WPK89}.

\begin{figure}[b]
  \centering
  \begin{minipage}{.5\textwidth}
    \centering
    \includegraphics[scale=0.32]{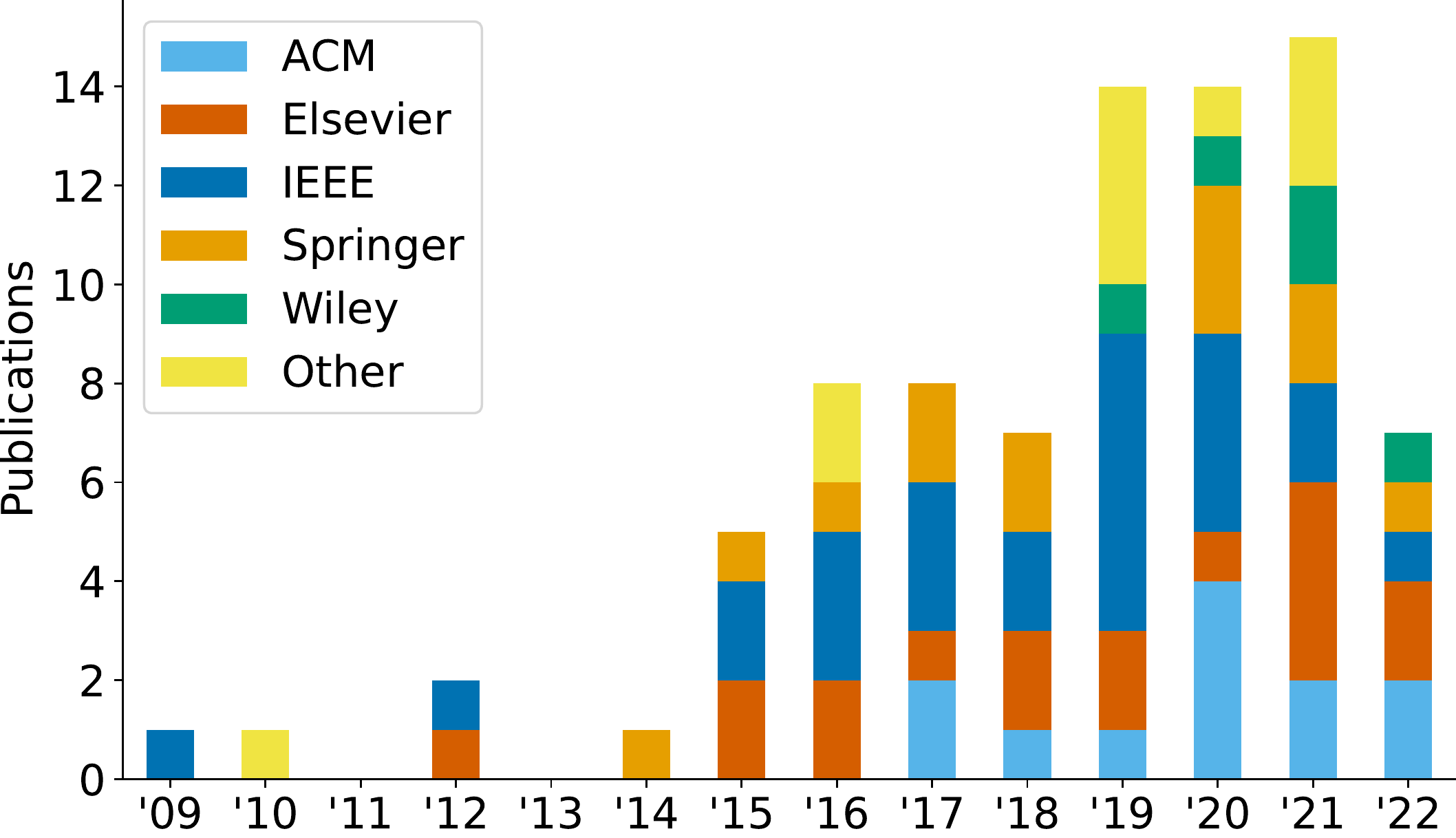}
    \captionof{figure}{Publications by year and publisher.}
    \label{fig:PubYear}
  \end{minipage}%
  \begin{minipage}{.5\textwidth}
    \centering
    \includegraphics[scale=0.32]{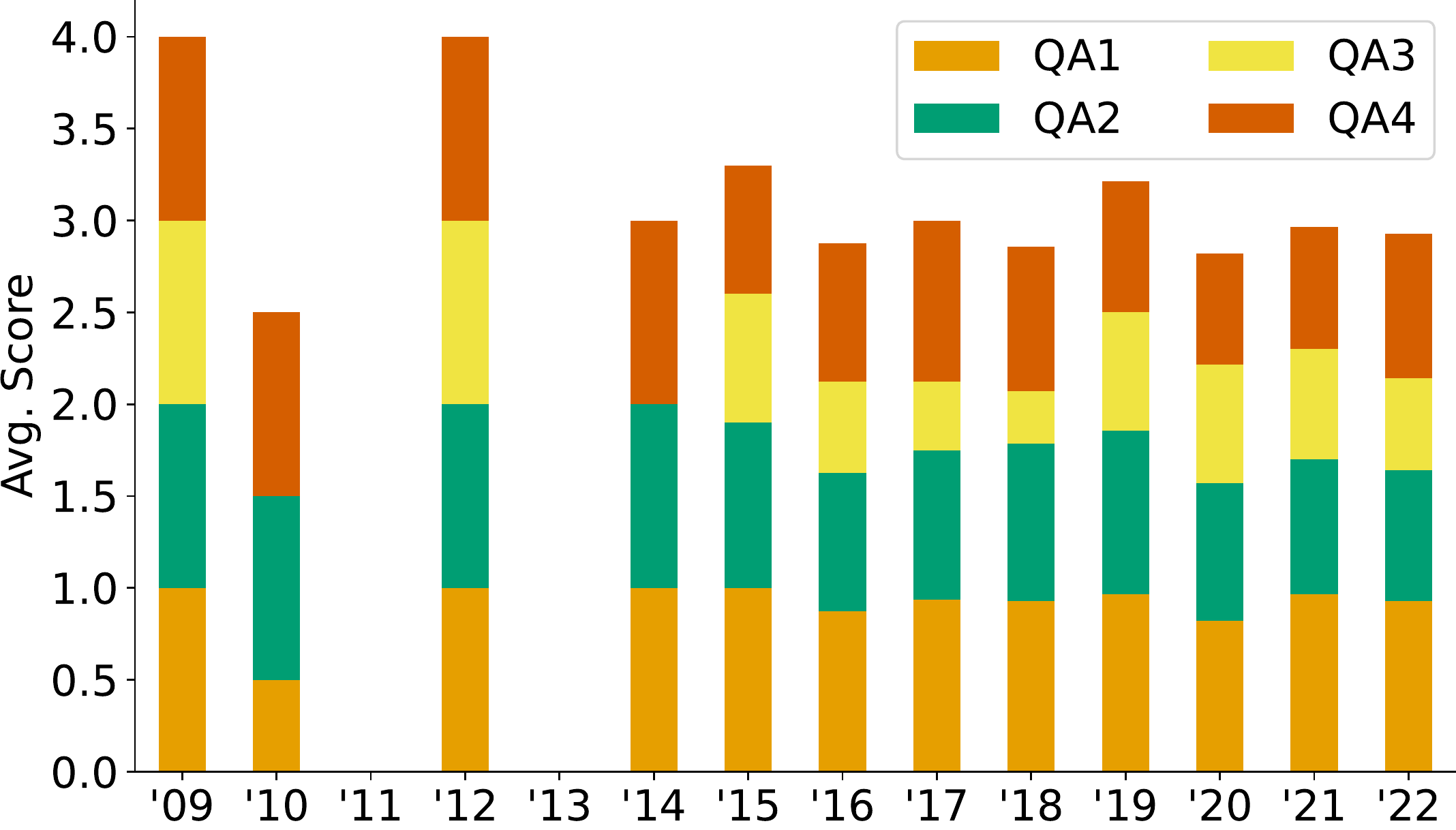}
    \captionof{figure}{Quality assessment scores by year and question.}
    \label{fig:QualityYear}
  \end{minipage}
\end{figure}

\subsection{RQ1: What SE tasks have been tackled with ML techniques?}
\label{sec:rq1}
The classification of the 83 studies according to the SWEBOK KAs and subareas,
as described in Section~\ref{sec:data-extraction},
revealed a coverage of the eleven KAs presented in Table~\ref{tab:RQ1tab}.
For each KA and subarea,
the number, percentage, and references of the associated secondary studies are included
as well as the number of primary studies reviewed by the secondary.
The most addressed KA is \emph{Software Quality},
being the focus of 25 (30\%) secondary studies.
Subsequent KAs include
\emph{Software Testing} ($n=$ 17; 20\%),
\emph{SE Process} ($n=$ 14; 18\%),
\emph{SE Management} ($n=$ 12; 14\%), and
\emph{Software Requirements} ($n=$ 6; 6\%).
A single study was found related to \emph{Engineering Foundations},
covering the subarea of \emph{Statistical Analysis},
despite this KA being less related to SE.
In Fig.~\ref{fig:AreasYear} we also visualize the yearly distribution of KAs.
KAs are sorted in the figure bottom-up
according to their appearance frequency in Table~\ref{tab:RQ1tab}.
It appears that \emph{Software Quality}, \emph{Software Testing}, and \emph{SE Process} are also
the most trending ones in recent years.
In the following sections we present for each KA,
the SE tasks that have been tackled with ML techniques.
Codes resulting from the manual coding process for the extraction of SE tasks
from the associated studies are set in \task{bold}.

\begin{table}[t]
  \caption{SWEBOK KAs, Subareas, and Primary (Prim.) Studies Covered by Secondary (Sec.) Studies}
  \label{tab:RQ1tab}
  \resizebox{\textwidth}{!}{
  \begin{tabular}{l l r r p{5.7cm} r}
    \hline
    \textbf{SWEBOK KA}
      & \textbf{Subarea} & \textbf{Sec.} & \textbf{\%} & \textbf{References} & \textbf{Prim.} \\
    \hline
    \multirow{4}{*}{Software (SW) Quality}
      & \multirow{2}{*}{Practical Considerations} & \multirow{2}{*}{22} & \multirow{2}{*}{27} &
      \cite{ML20, MB15, HTG19, AR19, TBA17, MB16, RMT09, AAG21, LM22, ECIA20, AAA20, PAKS22,
      APSW19, OT18, CS20, MC16, SUBG21, MAAK21, SPKK19, BFEA21, ECIA19, MK19} & \multirow{2}{*}{1\,309} \\
      & SW Quality Fundamentals & 2 & 2 & \cite{SS18, CRCP19} & 471 \\
      & SW Quality Management Processes & 1 & 1 & \cite{SB20} & 32 \\
    \hline
    \multirow{3}{*}{SW Testing}
      & \multirow{2}{*}{Test Techniques} & \multirow{2}{*}{15} & \multirow{2}{*}{18} &
      \cite{CGB20, ZSG16, OQCH19, KJMG17, AZAB17, RLJA18, Mal15, HBBG12, MDDC19, BK22,
      PMT21, SZWA20, KIJS21, AG20, RAJH21} & \multirow{2}{*}{1\,255} \\
      & Test Process & 2 & 2 & \cite{DDBE19, KBAA19} & 106 \\
    \hline
    \multirow{4}{*}{SE Process}
      & SW Life Cycles & 4 & 5 &
      \cite{SYA22, CA20, WCPM22, YXLB22} & 649 \\
      & SW Measurement & 4 & 5 & \cite{EDR18, MKR17, CTH16, AHS20} & 303 \\
      & \makecell[l]{SW Process Assessment \& Improvement} & 3 & 4 &
      \cite{SLLD16, DLLW17, SGG21} & 189 \\
      & SE Process Tools & 3 & 4 & \cite{ABDS18, YXLG21, LCB20} & 608 \\
    \hline
    SE Management
      & SW Project Planning & 12 & 14 &
      \cite{NZM19, PCCR20, KIA16, IHA16, WLLH12, Rad10, AG19a, MKAK22, SSB21,	ANA21,
      WSCW16, AG19b} & 563 \\
    \hline
    \multirow{3}{*}{SW Requirements}
      & Requirements Analysis & 2 & 2 & \cite{AY18, ZZA21} & 82 \\
      & Requirements Elicitation & 2 & 2 & \cite{BKS15, AFKK20} & 25 \\
      & Requirements Process & 2 & 2 & \cite{AHC21, AS15} & 60 \\
    \hline
    \multirow{3}{*}{SW Maintenance}
      & Techniques for Maintenance & 1 & 1 & \cite{LLN20} & 109 \\
      & Key Issues in SW Maintenance & 1 & 1 & \cite{UGDN17} & 32 \\
      & SW Maintenance Tools & 1 & 1 & \cite{TAB19} & 31 \\
    \hline
    \multirow{2}{*}{SW Design}
      & SW Structure \& Architecture & 1 & 1 & \cite{ZMI17} & 22 \\
      & SW Design Tools & 1 & 1 & \cite{LSS17} & 25 \\
    \hline
    \makecell[l]{SW Configuration Management}
      & \makecell[l]{SW Configuration Management Tools} & 1 & 1 & \cite{PAMJ21} & 69 \\
    \hline
    SE Models \& Methods
      & Analysis of Models & 1 & 1 & \cite{BRA14} & 79 \\
    \hline
    SE Professional Practice
      & Group Dynamics \& Psychology & 1 & 1 & \cite{GFS21} & 63 \\
    \hline
    Engineering Foundations
      & Statistical Analysis & 1 & 1 & \cite{IAA15} & 35 \\
    \hline
  \end{tabular}}
\end{table}

\begin{figure}[t]
  \centering
  \includegraphics[scale=0.5]{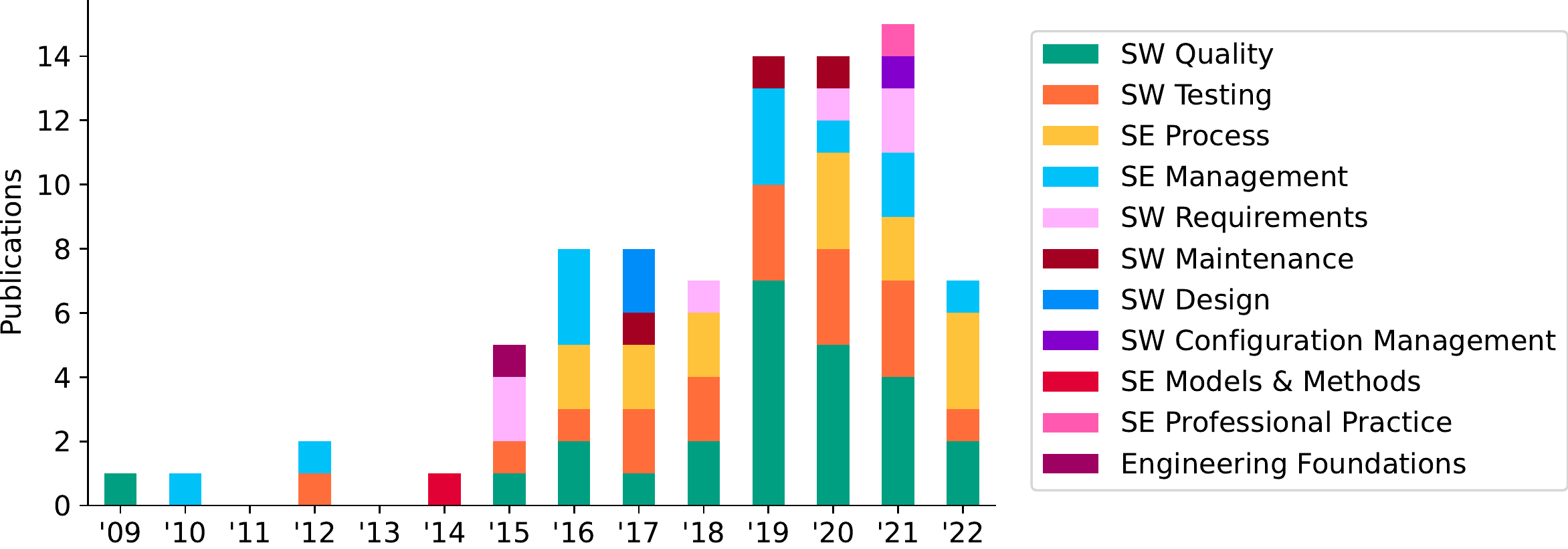}
  \caption{Publications by year and knowledge area.}
  \label{fig:AreasYear}
\end{figure}

\subsubsection{Software Quality}
In this area ML techniques are frequently used for
\task{software change} and \task{quality prediction},
replacing traditional statistical methods.
ML-based software change prediction aims to identify change-prone components
during the early phases of software development,
leading to higher quality and maintainable software at lower cost~\cite{MB15,MB16}.
To evaluate ML techniques' effectiveness,
Malhotra and Khanna~\cite{MK19} summarize different software change prediction models,
experimental setups,
data analysis algorithms,
statistical validation tests, and associated threats.
Feature selection is considered one of the most essential and complex activities
in data pre-processing~\cite{AR19}.
In software quality prediction
we identify various applications of Bayesian networks,
while the majority of the reviewed work employs expert knowledge~\cite{TBA17}.

Another common task in software quality is \task{software maintainability prediction}.
By measuring quality attributes,
developers can improve software design,
optimize resource allocation, and
develop cost-effective, high-quality, maintainable software systems~\cite{MC16,ML20}.
Managers can evaluate and compare productivity across projects,
perform effective resource planning, and
control maintenance costs~\cite{MC16}.
An accuracy analysis of software maintainability prediction models reveals that
the most accurate ones are
fuzzy and neuro fuzzy systems, and artificial neural networks~\cite{ECIA19}.
Moreover,
the use of open-source datasets in the training process
of maintainability prediction models has increased in recent years~\cite{MC16}.

ML-based \task{software defect prediction} is widely covered in software quality.
Uncovering defects at the early phases of software development
improves the system's cost-effectiveness and maintainability~\cite{OT18}.
Some reviews recommend the use of supervised and
deep learning techniques~\cite{MAAK21,BFEA21}.
Pachouly \etal~\cite{PAKS22} propose an architecture for synthesizing training datasets
for multi-label classification of software defects.
A popular topic is \emph{cross project defect prediction}:
training ML models with data gathered from various projects
to predict defects in completely new (unseen) projects~\cite{HTG19}.
An \emph{ideal} prediction model should
\emph{highlight the severity of defects,
uncover security-related defects and system vulnerabilities, and
also identify defects on systems that it was not trained on}~\cite{SPKK19}.

Other popular quality tasks include \task{malware}, \task{smell}, and
\task{data exfiltration detection}.
In JavaScript
ML models offer higher malware detection rates than non-ML ones~\cite{SB20}.
In smell detection
we distinguish the use of deep learning,
J48, JRip, Decision Tree, and Random Forest algorithms~\cite{AAA20,APSW19,CRCP19,LM22}.
These involve training models
on datasets of source code metrics and smells
to classify new unseen software components~\cite{SS18}.
Some efforts target smell detection at the design and code levels~\cite{AAG21}.
In the area of data exfiltration
(also known as \emph{data theft} and \emph{data leakage}~\cite{UERC18})
there are several ML-based countermeasures,
which are classified according to four axes:
type of approach (data-- or behavior-driven);
employed features (behavioral, content-based, statistical, syntactical, spatial,
or temporal);
evaluation dataset (simulated, synthesized, or real); and
reported performance measures~\cite{SUBG21}.

\subsubsection{Software Testing}
\label{sec:rq1-testing}
Here
ML techniques are mainly used for \task{test automation},
specifically for test case generation, evaluation, and optimization.
Genetic algorithms are frequently proposed~\cite{OQCH19,KBAA19}.
In the test oracle problem,
which is concerned with a software's output behavior based on a set of inputs,
automation is achieved by training ML models to predict the outcome~\cite{DDBE19}.
In exhaustive testing,
where developers typically use auto-generated test data,
ML algorithms are used to produce ample efficient test case routines~\cite{DDBE19}.
In mobile application testing the Swift Hand automated technique~\cite{CNS13}
can be used to produce sequences of test inputs
that enable visiting unexplored states of
the application without restarting it~\cite{ZSG16}.
In test evaluation
testers assess with ML models the extent to which
test suites cover the observable program behavior, and
predict the feasibility of test cases~\cite{DDBE19}.
In test optimization
historical information is analyzed with ML techniques to
calculate multi-objective metrics~\cite{KJMG17}.
These metrics are used to improve testing performance
by reducing resource and time consumption,
especially in regression testing where retesting is costly~\cite{KIJS21,RAJH21}.
ML-based test case prioritization is useful
when the source code is not accessible, and
for enhancing fault detection in applications~\cite{CGB20}.
Lastly,
Black-box test specification can be refined with ML techniques~\cite{RLJA18}.

Furthermore,
ML is frequently used in \task{software fault prediction}.
In large datasets
deep learning models outperform data mining and ML ones~\cite{BK22}.
In interaction testing
ML techniques are used to create diverse test suites,
covering as many constraints as possible~\cite{AZAB17}.
ML-based classification algorithms effectively determine a module
or class's fault-proneness,
usually more efficiently than statistical models~\cite{Mal15,PMT21}.
The results of ML methods are also more generalizable
than statistical ones~\cite{PMT21}.
Still,
ML techniques can perform worse than statistical models,
because they require parameter fine-tuning,
and may depend on unbalanced or uncleaned data~\cite{HBBG12}.
To perform parameter optimization and feature selection,
bio-inspired algorithms are recommended,
namely genetic algorithms and particle swarm optimization~\cite{AG20}.

Another testing area of ML interest is \task{software vulnerability detection}.
Deep learning methods (convolutional and recurrent neural networks)
are frequently employed for vulnerability analysis on source code~\cite{SZWA20}.
The majority of training datasets consist of data
from a single programming language,
mainly C/C++ and JavaScript~\cite{SZWA20}.
In penetration testing
ML applications mainly involve attack planning
through attack graph generation or attack tree modeling~\cite{MDDC19}.
Two predominant methods for attack planning are
the Markov Decision Process and genetic algorithms~\cite{MDDC19}.

\subsubsection{SE Process}
Various tasks related to SE processes are approached with ML.
Some \task{software fault prediction} models are developed based on topic modeling.
Topic modeling is applied on source code to
approximate software concerns as topics,
analyze failed executions and the defect-proneness of systems, or
determine dependencies between source code elements and developers~\cite{SLLD16}.
Apart from fault prediction,
other applications of topic modeling are:
architecting, bug handling, coding, documentation,
maintenance, refactoring, requirements, and testing~\cite{SGG21}.
In \task{software quality prediction}
ML techniques are employed to estimate four predominant quality attributes:
effort, defect-- and change-proneness, and maintainability~\cite{MKR17}.
Furthermore,
ML models are preferred for
\task{traceability link recovery},
\task{concept location},
\task{bug triaging},
\task{clone identification},
\task{source code summarization}, and
\task{refactoring}~\cite{CTH16,AHS20}.

A number of reviews concern \task{software process mining} and \task{automation}.
In agile software development,
process mining is used to discover processes followed by agile teams
based on task tracking applications~\cite{EDR18}.
The most prominent tool for this purpose is ProM~\cite{DMVW05}.
In software effort estimation,
process mining is used to improve the accuracy of models~\cite{DLLW17}.
In general,
ML is intensively used in the requirements phase
for automation and improvement~\cite{SYA22}, and
in the maintenance phase for prediction~\cite{YXLB22}.
It is also widely used for the automation of decision-making tasks
and for predictive analysis~\cite{SYA22,YXLB22}.
Hybrid models combining ML with AI techniques improve the models' performance,
notably when training datasets consist of
multi-dimensional and diverse instances~\cite{SYA22}.
Furthermore,
deep learning techniques are increasingly employed in SE processes
due to their automated feature engineering capabilities,
superior efficiency, and
ability to replace human expertise~\cite{YXLB22,WCPM22}.

Other emerging areas of SE process are \task{source code analysis},
\task{program generation}, and \task{recommender systems}.
Source code is contrasted to natural language
since their common statistical properties can be used
to build better SE tools~\cite{ABDS18}.
Code-generating, code representational, and pattern-mining
ML and deep learning models find applications in:
recommender systems;
coding convention inference;
defect detection;
code translation, copying, and cloning;
code-to-text and text-to-code;
documentation, traceability, and information retrieval; and
program synthesis and analysis~\cite{ABDS18,YXLG21,LCB20}.

\subsubsection{SE Management}
In this area we distinguish ML-based \task{software development cost}
and \task{effort estimation}.
Decision trees preponderate in software development cost estimation~\cite{NZM19}, and
use case points stand out in software development effort prediction~\cite{ANA21}.
Ensemble effort estimation models,
which are  usually constructed from single (\ie base) models
joined with linear and non-linear combiner functions,
perform better than single techniques~\cite{MKAK22,IHA16}.
In general,
the estimation accuracy of ML models is superior than that of regression models,
although the application of the former in the industry
is still limited~\cite{PCCR20,KIA16,WLLH12}.
Again,
bio-inspired feature selection algorithms improve even further
the accuracy of ML models~\cite{AG19b}.
Bayesian networks can also introduce expert knowledge in the models
through strictly-defined probability functions,
especially when no empirical data from older
projects are available~\cite{Rad10}.

A small number of reviews concern
\task{software enhancement (maintenance) effort estimation}~\cite{SSB21}.
Regression problems of enhancement effort prediction are addressed
with single prediction models,
which outperform statistical ones~\cite{SSB21}.
In open source software,
effort estimation and maintenance activity time prediction (\eg bug fixing) models
are trained on source code and people-related metrics~\cite{WSCW16}.

\subsubsection{Software Requirements}
Here ML is used to support a range of tasks including:
\task{volatility prediction}~\cite{AY18};
\task{reuse of requirements} and \task{feature and variability extraction}
in the context of software product lines~\cite{BKS15};
\task{requirements elicitation}, \task{analysis}, \task{specification}, \task{prioritization}, and
\task{negotiation}~\cite{AHC21,AS15,AFKK20}; and
\task{requirements ambiguity resolution}~\cite{ZZA21}.

\subsubsection{Software Maintenance}
\task{Bug prioritization} and \task{software rename refactoring}
are frequently approached with ML techniques in this area.
Especially natural language processing models
are used to mine, summarize, and prioritize bug reports~\cite{UGDN17,TAB19}.
In software rename refactoring
ML techniques outperform traditional approaches
that are based on empirical rules~\cite{LLN20}.

\subsubsection{Software Design}
Researchers here apply ML to \task{software architecture recovery},
\task{reverse engineering variability}, and
\task{feature and variability extraction}.
In software architecture recovery,
ML-based classification techniques are recommended
for the estimation of the positional probability of edges of a
weighted graph using information gathered from real-world object-oriented reference software systems~\cite{ZMI17}.
Reverse engineering variability and feature and variability extraction
mainly concern the process of
migrating individual software systems to software product lines~\cite{LSS17}.

\subsubsection{Software Configuration Management}
In this area we distinguish efforts in \task{software performance prediction} and
\task{software configuration interpretability and optimization}.
The large number of configuration options of modern software systems makes it impossible
to explore the entire configuration space,
to satisfy functional and non-functional needs~\cite{PAMJ21}.
As a workaround,
researchers frame the software configuration problem as a ML one
by training models on samples of configuration measurements~\cite{PAMJ21}.
The software configuration problem is also related to
compiler autotuning and database management system tuning~\cite{PAMJ21}.

\subsubsection{SE Models and Methods}
ML techniques are proposed for
systematically retrieving large-scale information from software artifacts
to support \task{trace recovery}~\cite{BRA14}.
Specifically,
trace link structures are extracted from software artifacts, and
are used as input to ML voting systems
to assess the importance of each artifact~\cite{BRA14}.

\subsubsection{SE Professional Practice}
Here ML techniques are used to \task{estimate the cognitive load}
of software developers when performing SE tasks.
To improve estimation results,
these techniques are usually combined with feature sets
related to developers' cognitive load~\cite{GFS21}.

\subsubsection{Engineering Foundations}
In this area
ML is used in the task of \task{missing value imputation} in SE datasets,
to improve the accuracy
of software development effort estimation results~\cite{IAA15}.

\subsection{RQ2: What SE knowledge areas could be better covered by ML techniques?}
\label{sec:rq2}
From Table~\ref{tab:RQ1tab}
we deduce that some SWEBOK KAs are either covered
by few ML-related secondary studies,
or are not covered at all.
Specifically,
\emph{Software Construction}
and \emph{SE Economics} are not addressed by any secondary study.
Moreover, \emph{Software Configuration Management},
\emph{SE Models and Methods}, and
\emph{SE Professional Practice} are only
addressed by a single secondary study each,
therefore, they appear less covered compared to the other KAs.
Some KAs are also more comprehensively covered than others
as their associated studies span more subareas.
This is the case for \emph{Software Quality},
\emph{SE Process},
\emph{Software Requirements}, and
\emph{Software Maintenance},
which span various subareas,
as opposed to \emph{SE Management},
which contains an equivalent number of reviews mapped to a single subarea.
The sparse coverage of certain KAs is also recognized
by the authors of many secondary studies
through their calls for further research on the application of ML techniques
to the associated SE tasks.
In the subsequent sections
we provide an extensive description of how each KA and SE task could be better covered,
as evidenced from the authors' remarks
through the process described in Section~\ref{sec:data-extraction},
as well as any issues and obstacles related to the use of ML techniques in SE.
We also provide some general recommendations that apply to all KAs
in Section~\ref{sec:rq2-general}.

\subsubsection{General Recommendations}
\label{sec:rq2-general}
Various recommendations apply to all KAs.
These include:
conducting more comparative analyses between ML methods
and traditional statistical techniques ($n=$ 13 studies);\footnote{The complete list
of associated studies for each general recommendation is available
in file \emph{further\_research\_general.txt}.}
performing more empirical analyses to increase the quality of ML techniques,
and establish their cost-effectiveness ($n=$ 12);
publishing and using more open-source, diverse, large-scale, and
high-quality datasets ($n=$ 16);
experimenting with new, hybrid, ensemble, and incremental techniques
(\eg transfer, few-shot, weakly, semi-supervised, and active learning,
blockchain technology,
search-based and multi-objective methods,
automated feature engineering techniques,
regression-based methods---$n=$ 21);
ensuring industrial relevance and scalability of ML models by applying them
in real settings and commercial datasets,
by conducting in-depth case studies with practitioners, and
by developing concrete methods for building reliable models ($n=$ 18);
optimizing hyper-parameters of ML models
(\eg with the grid search algorithm---$n=$ 3); and
handling class imbalance in training datasets and model over-fitting ($n=$ 3).

\subsubsection{Software Quality}
A number of research suggestions are observed in the associated SE tasks of this area.
In software change and quality prediction
researchers should explore cross-organization and cross-company validation,
effective transfer learning, and temporal validation~\cite{MK19}.
In software maintainability prediction
we observe literature ambiguities concerning maintainability definitions and characteristics
that affect ML models' performance~\cite{ECIA20}.
Researchers propose:
measuring design metrics dynamically instead of statically
(\eg with Dynamic Lack of Cohesion, Dynamic Response For a Class)~\cite{MC16};
investigating maintainability for other types of applications
(\eg web, mobile, model-driven, and cloud computing)~\cite{MC16,ECIA19};
analyzing the effect of diverse software development and process management factors
on maintainability
(\eg risk analysis, project planning, requirement analysis)~\cite{MC16,ECIA19}; and
focusing on maintainability before delivery of the software product
to detect issues and quality failures early~\cite{ECIA19}.
In software defect prediction
there are needed more empirical analyses on the validity
of cross project defect prediction datasets~\cite{HTG19,PAKS22,SPKK19}.
Furthermore,
in smell detection further research is needed on:
smell types~\cite{LM22};
smell prioritization~\cite{AAG21};
the mutual effect between smells to identify highly correlated ones~\cite{AAG21,APSW19};
the false-positiveness of the proposed ML techniques,
which could be high due to deficient smell definitions~\cite{SS18,LLN20,AAG21,AAA20}; and
the effect of data transformations on the smell detection process~\cite{AAG21}.

\subsubsection{Software Testing}
Here we summarize the following promising research directions.
In ML-based mobile application testing
most existing studies only target the Android platform,
therefore there is a need for including other prevalent platforms
as well (\eg Apple iOS, Windows Phone)~\cite{ZSG16}.
In mutation testing
there is considerable room for expediting existing ML-based solutions
for detecting possible equivalent mutants
as well as automating more facets of the process,
which are currently handled by humans~\cite{DDBE19}.
In test suite optimization
there are only few ML-based approaches~\cite{KBAA19}.
In test case prioritization,
including software requirement attributes in the training datasets
could possibly improve models' effectiveness~\cite{RAJH21}.
In software fault prediction
researchers could investigate less popular ML and evolutionary algorithms,
such as Logit Boost, Ada Boost, Rule Based Learning, Bagging,
Alternating Decision Tree, Radial Basis Function,
Ant Colony Optimization, and Genetic Programming~\cite{PMT21,AZAB17}.
More empirical analyses are desired on the performance of
bio-inspired parameter optimization and feature selection algorithms
with the traditional algorithms
(\eg grid search, random search, greedy search, best-first search)~\cite{AG20}.
Moreover,
in software vulnerability prediction
assessment criteria should include more qualitative measures (\eg consequence, impact)
for determining the effectiveness  of ML techniques~\cite{MDDC19}.
Suggestions for ML models include:
designing approaches to identify exploitable vulnerabilities in real-time~\cite{MDDC19};
studying the impact of multiple programming language datasets~\cite{SZWA20}; and
performing finer-grained vulnerability detection
by identifying the precise location of the vulnerabilities
in the source code files or functions~\cite{SZWA20}.

\subsubsection{SE Process}
Regarding SE processes,
in the tasks of software fault and quality prediction, and
traceability link recovery we recognize the following recommendations.
In fault and quality prediction
researchers should further explore and assess the employed search-based techniques
through multiple evaluation factors including
cost-effectiveness,
comprehensibility,
generalizability, and
execution time~\cite{MKR17}.
To ease the application of search-based approaches in prediction models,
practitioners could develop and establish relevant tool suites~\cite{MKR17}.
To reduce bias in the evaluation process,
different validation techniques such as
inter-release,
cross-project, and
temporal validation could be combined~\cite{MKR17}.
Subsequent research should thoroughly document
its associated threats to validity~\cite{MKR17}.
In traceability link recovery more emphasis should be placed on:
recovering links between trace artifacts
that are commonly used in modern software development
(\eg user stories, accepted test cases and source code);
building traceability systems beyond text-based recovery
(\eg recovering traceability links between design images and requirements);
advanced static program analysis,
such as value flow analysis~\cite{SX16} and pointer aliasing analysis~\cite{LS19},
to support more precise change impact analysis; and
evaluating industrial datasets and survey practitioners
to gain valuable feedback for further improvements~\cite{AHS20}.

In software process mining and automation the following remarks are observed.
There is a need for:
more approaches for extracting requirement-related information
from community forums;
replicable, standardized, baseline approaches for comparisons;
experiments with transformers (\eg BERT) in clone detection and program repair;
exploring less popular pre-processing techniques
(\eg directed graphs, lookup tables, execution trace vectors);
increasing the interpretability of deep learning solutions;
developing guidelines and a supporting infrastructure
for comparing metrics and evaluating ML approaches;
standardizing the data pre-processing pipelines to reduce bias; and
more user studies to gain insights into when and where
automated techniques are useful to humans and more accurate
than manual methods~\cite{SYA22,WCPM22,YXLB22}.

Numerous suggestions are made in the tasks of
source code analysis, program generation, and recommender systems.
Some promising areas for deep learning applications are:
debugging, traceability, code completion and synthesis, education, and
assistive tools (\eg IDEs)~\cite{ABDS18}.
Researchers propose:
building web platforms associated with the ML models;
developing modular neural network architectures
to combat issues with compositionality, sparsity, and generalization;
developing deep learning models that fit multiple programming languages;
more concise and discrete representations of language and code
to perform complex reasoning and predictions;
constructing an evaluation metric that can incorporate
both semantic meaning and grammatical and execution correctness;
simplifying deep source code models by learning from human experience;
employing generative models to address the representation learning issues
in program generation;
developing more practical and engineering-friendly frameworks
for rapid new concept learning and code generation; and
experimenting with code-in-code-out systems for generation
and evaluation~\cite{SYA22,ABDS18,LCB20}.

Various ideas inferred from the secondary reviews of SE process concern
further experimentation with topic modeling.
Fruitful application areas include:
feature location;
software regression testing;
developer recommendation;
software refactoring;
software fault prediction;
traceability link recovery; and
software analytics~\cite{SLLD16,MKAK22,SGG21,CTH16}.
Suggested applications of topic modeling are:
searching collections of software systems;
measuring the evolutionary trends of repositories;
establishing traceability links between emails and source code; and
analyzing software systems by applying topic models on email archives
and execution logs~\cite{CTH16}.
To improve the results of prior studies,
replications could be conducted after
fine-tuning the parameters of topic models, and
improving the data pre-processing
by analyzing the value of query expansion and context consideration~\cite{CTH16}.
Future topic models could incorporate the structure of software development data~\cite{CTH16}.

\subsubsection{SE Management}
Here we distinguish the following recommendations related
to software effort and cost estimation.
More empirical research is needed on:
the application of rarely-used ML techniques,
in order to help researchers formulate better processes,
and assist practitioners in decision making~\cite{WLLH12,PCCR20,KIA16};
the performance evaluation of ensemble estimation techniques
that are based on regression trees and case-based reasoning~\cite{IHA16,PCCR20}; and
the accuracy assessment of bio-inspired feature selection algorithms~\cite{AG19b}.
In addition,
more experiments should be conducted with:
heterogeneous ensemble effort estimation models
(\ie models that combine at least two different base models);
ML models that employ genetic programming and genetic algorithms;
cascade correlation neural networks;
developer-related metrics (\eg individual contribution and performance)
to predict bug fixing time; and
size-related metrics
to estimate open source software maintenance effort~\cite{IHA16,AG19a,WSCW16}.

\subsubsection{Software Requirements}
The following research gaps are observed in this area.
Further empirical research is needed
on how to assess and select the most suitable ML techniques
in requirements volatility prediction and ambiguity resolution~\cite{AY18,ZZA21}, and
how to automate with ML the extraction of software requirements
from natural language documents~\cite{BKS15,AHC21}.
Researchers should experiment with ML in more requirements activities,
such as requirements specifications and management~\cite{AHC21}.
Some research ideas include
sharing standard pre-labeled datasets,
standardizing nonfunctional requirements,
applying sentiment analysis on functional and nonfunctional requirements, and
performing change impact analysis in ML-based software requirements~\cite{AFKK20,ZZA21}.

\subsubsection{Software Maintenance}
Here we distinguish recommendations in bug prioritization and software rename refactoring.
In bug prioritization
researchers suggest further exploring ML and
bug tossing (\ie reassignment) graphs~\cite{JKZ09}
for automating developer assignment in bug reports~\cite{UGDN17} as well as
exploiting ML to automate the process of bug report summarization~\cite{TAB19}.
In software rename refactoring
there is a need for
more models that automatically execute and detect renamings
of software entities~\cite{LLN20}.
Suggestions include investigating
the usefulness of advanced identifier splitting approaches,
the preservation of name bindings based on language features, and
the performance of different renaming representation techniques~\cite{LLN20}.
Although the primary purpose of renaming is to improve program comprehension,
researchers should also consider its reverse application:
using ML-based renaming techniques for identifier obfuscation~\cite{LLN20}.

\subsubsection{Software Design}
There is fertile ground in software architecture recovery.
Future studies should go beyond recovering components and connectors
of software architectures,
to identifying
the employed design patterns and architectural styles, as well as
the associated system concerns in existing software systems~\cite{ZMI17}.
More analyses are suggested on recovered architectures
with respect to their conceivable similarity with the legacy systems' architecture~\cite{ZMI17}.
A prospective direction could be
identifying faults in recovered architectures,
that could possibly lead to system failures
either during or after the maintenance of the legacy systems~\cite{ZMI17}.

\subsubsection{Software Configuration Management}
In software performance prediction, configuration interpretability, and optimization,
although the ML models' results are quite accurate,
there is still room for reducing learning errors, and 
generalizing predictions to multiple computing environments~\cite{PAMJ21}. 

\subsubsection{SE Models and Methods}
To improve trace recovery,
Borg \etal~\cite{BRA14} suggest
combining probabilistic retrieval methods with ML techniques, and
conducting more research on the scalability of ML models in large projects.

\subsubsection{SE Professional Practice}
In cognitive load estimation the following remarks are noted.
SE research currently lacks effective tools and methods
for measuring and evaluating practitioners' cognitive load~\cite{GFS21}.
For this,
more replication studies are needed that document in detail
any experiences and lessons learned from the application of ML-based methods
for cognitive load estimation~\cite{GFS21}.
Two promising research directions concern
evaluating the effectiveness of psychophysiological metrics
in cognitive load estimation, and
predicting the unproductive periods of developers,
\eg identifying when their cognitive load levels are error-- and bug-prone.~\cite{GFS21}.

\subsubsection{Engineering Foundations}
In the task of missing value imputation
researchers could experiment with ML techniques
that have not been employed yet in the SE field,
including novel ideas and methods from related disciplines~\cite{IAA15}.

\subsection{RQ3: What ML techniques have been used in SE?}
\label{sec:rq3}
The classification of the 83 studies according to the four axes described
in Section~\ref{sec:data-extraction} is presented in Table~\ref{tab:MLclass}.
With regard to the \emph{role of AI in SE},
the majority of studies ($n=$ 54; 65\%) were classified in the
\emph{Classification, learning and prediction} category, followed by
\emph{Fuzzy and probabilistic methods for reasoning in the presence of uncertainty} ($n=$ 17; 20\%)
and \emph{Computational search and optimisation techniques} ($n=$ 12; 14\%).
In the \emph{supervision} axis,
most studies ($n=$ 65; 78\%) adopt supervised learning, followed by
unsupervised ($n=$ 11; 13\%),
semi-supervised ($n=$ 5; 6\%), and
reinforcement learning ($n=$ 2; 2\%).
According to the \emph{incrementality} axis,
almost all studies perform batch/offline learning,
and only one study appears to adopt online/incremental learning.
Finally, in the \emph{generalizability} axis,
the majority of studies ($n=$ 72; 87\%) perform model-based learning,
while the remaining ($n=$ 11; 13\%) employ instance-based learning.

\begin{table}[t]
  \caption{ML Techniques Employed by Secondary Studies}
  \label{tab:MLclass}
  \resizebox{\textwidth}{!}{
  \begin{tabular}{c p{7cm} r r p{9cm}}
    \hline
    \textbf{Axis}	& \textbf{Technique}	& \textbf{Total}	& \textbf{\%}	& \textbf{Studies} \\
    \hline
    \multirow{5}{*}{{\rotatebox[origin=c]{90}{{\shortstack[c]{Role of AI \\ in SE}}}}}
      & Computational search and optimisation techniques
        & 12	& 14
          & \cite{OQCH19, EDR18, KJMG17, MKR17, AZAB17,
                  KBAA19, AAG21, MDDC19, AG20, AG19b,
                  CRCP19, RAJH21} \\
      & \multirow{2}{*}{\shortstack[l]{Fuzzy and probabilistic methods for reasoning in the \\ presence of uncertainty}}
        & 17	& 20
          & \cite{SLLD16, AY18, TBA17, KIA16, CTH16,
                  IHA16, WLLH12, Rad10, BRA14, LSS17,
                  BKS15, SGG21, PAMJ21, AHS20, AFKK20,
                  ABDS18, LCB20} \\
      & \multirow{3}{*}{Classification, learning and prediction}
        & \multirow{3}{*}{54}	& \multirow{3}{*}{65}
          & \cite{DDBE19, CGB20, IAA15, ZSG16, ZMI17,
                  NZM19, ML20, MB15, PCCR20, RLJA18,
                  HTG19, UGDN17, TAB19, AR19, GFS21,
                  SS18, Mal15, LLN20, MB16, RMT09,
                  DLLW17, HBBG12, AG19a, AHC21, MKAK22,
                  LM22, AS15, ECIA20, AAA20, SSB21,
                  BK22, PAKS22, PMT21, APSW19, OT18,
                  ANA21, SZWA20, SB20, KIJS21, SYA22,
                  WSCW16, ZZA21, CS20, CA20, MC16,
                  SUBG21, WCPM22, YXLB22, MAAK21, SPKK19,
                  BFEA21, ECIA19, MK19, YXLG21} \\
    \hline
    \multirow{6}{*}{{\rotatebox[origin=c]{90}{Supervision}}}
      & \multirow{3}{*}{Supervised learning}
        & \multirow{3}{*}{65}	& \multirow{3}{*}{78}
          & \cite{DDBE19, CGB20, IAA15, ZSG16, NZM19,
                  OQCH19, ML20, EDR18, KJMG17, MKR17,
                  AZAB17, MB15, PCCR20, HTG19, UGDN17,
                  AR19, GFS21, AY18, SS18, KIA16,
                  LLN20, MB16, RMT09, DLLW17, WLLH12,
                  Rad10, HBBG12, AG19a, AAG21, AHC21,
                  MKAK22, LM22, AS15, ECIA20, AAA20,
                  BK22, PAKS22, PMT21, APSW19, OT18,
                  PAMJ21, SZWA20, SB20, KIJS21, SYA22,
                  AG20, WSCW16, ZZA21, AG19b, CS20,
                  CA20, MC16, CRCP19, SUBG21, WCPM22,
                  YXLB22, RAJH21, MAAK21, SPKK19, BFEA21,
                  ECIA19, MK19, ABDS18, YXLG21, LCB20} \\
      & Unsupervised learning
        & 11	& 13
          & \cite{ZMI17, SLLD16, RLJA18, TAB19, CTH16,
                  BRA14, LSS17, BKS15, SGG21, AHS20,
                  AFKK20} \\
      & Semi-supervised learning
        & 5	& 6
          & \cite{TBA17, Mal15, IHA16, SSB21, ANA21} \\
      & Reinforcement learning
        & 2	& 2
          & \cite{KBAA19, MDDC19} \\
    \hline
    \multirow{5}{*}{{\rotatebox[origin=c]{90}{{\shortstack[c]{Incremen-\\tality}}}}}
      & \multirow{4}{*}{Batch/offline learning}
        & \multirow{4}{*}{82}	& \multirow{4}{*}{99}
          & \cite{DDBE19, CGB20, IAA15, NZM19, OQCH19,
                  ML20, EDR18, KJMG17, MKR17, AZAB17,
                  MB15, PCCR20, HTG19, UGDN17, AR19,
                  GFS21, AY18, SS18, KIA16, LLN20,
                  MB16, RMT09, DLLW17, WLLH12, Rad10,
                  HBBG12, ZMI17, SLLD16, RLJA18, TAB19,
                  CTH16, BRA14, LSS17, BKS15, TBA17,
                  Mal15, IHA16, KBAA19, AG19a, AAG21,
                  AHC21, MKAK22, LM22, AS15, SGG21,
                  ECIA20, AAA20, SSB21, MDDC19, BK22,
                  PAKS22, PMT21, APSW19, OT18, PAMJ21,
                  ANA21, SZWA20, SB20, KIJS21, SYA22,
                  AG20, WSCW16, ZZA21, AG19b, CS20,
                  CA20, MC16, CRCP19, AHS20, SUBG21,
                  WCPM22, YXLB22, AFKK20, RAJH21, MAAK21,
                  SPKK19, BFEA21, ECIA19, MK19, ABDS18,
                  YXLG21, LCB20} \\
      & Online/incremental learning
        & 1	& 1
          & \cite{ZSG16} \\
    \hline
    \multirow{4}{*}{{\rotatebox[origin=c]{90}{{{\shortstack[c]{Generali-\\zability}}}}}}
      & \multirow{3}{*}{Model-based learning}
        & \multirow{3}{*}{72}	& \multirow{3}{*}{87}
          & \cite{DDBE19, CGB20, NZM19, OQCH19, ML20,
                  EDR18, KJMG17, MKR17, AZAB17, MB15,
                  PCCR20, HTG19, UGDN17, AR19, GFS21,
                  AY18, SS18, KIA16, LLN20, MB16,
                  RMT09, DLLW17, WLLH12, Rad10, HBBG12,
                  ZMI17, RLJA18, TAB19, TBA17, Mal15,
                  IHA16, KBAA19, AG19a, AAG21, AHC21,
                  LM22, AS15, ECIA20, AAA20, SSB21,
                  MDDC19, BK22, PAKS22, PMT21, APSW19,
                  OT18, PAMJ21, ANA21, SZWA20, SB20,
                  KIJS21, SYA22, AG20, WSCW16, ZZA21,
                  AG19b, CS20, CA20, MC16, CRCP19,
                  SUBG21, WCPM22, YXLB22, RAJH21, MAAK21,
                  SPKK19, BFEA21, ECIA19, MK19, ABDS18,
                  YXLG21, LCB20} \\
      & Instance-based learning
        & 11	& 13
          & \cite{IAA15, ZSG16, SLLD16, CTH16, BRA14,
                  LSS17, BKS15, MKAK22, SGG21, AHS20,
                  AFKK20} \\
    \hline
  \end{tabular}}
\end{table}

\begin{figure}[b]
  \centering
  \includegraphics[scale=0.5]{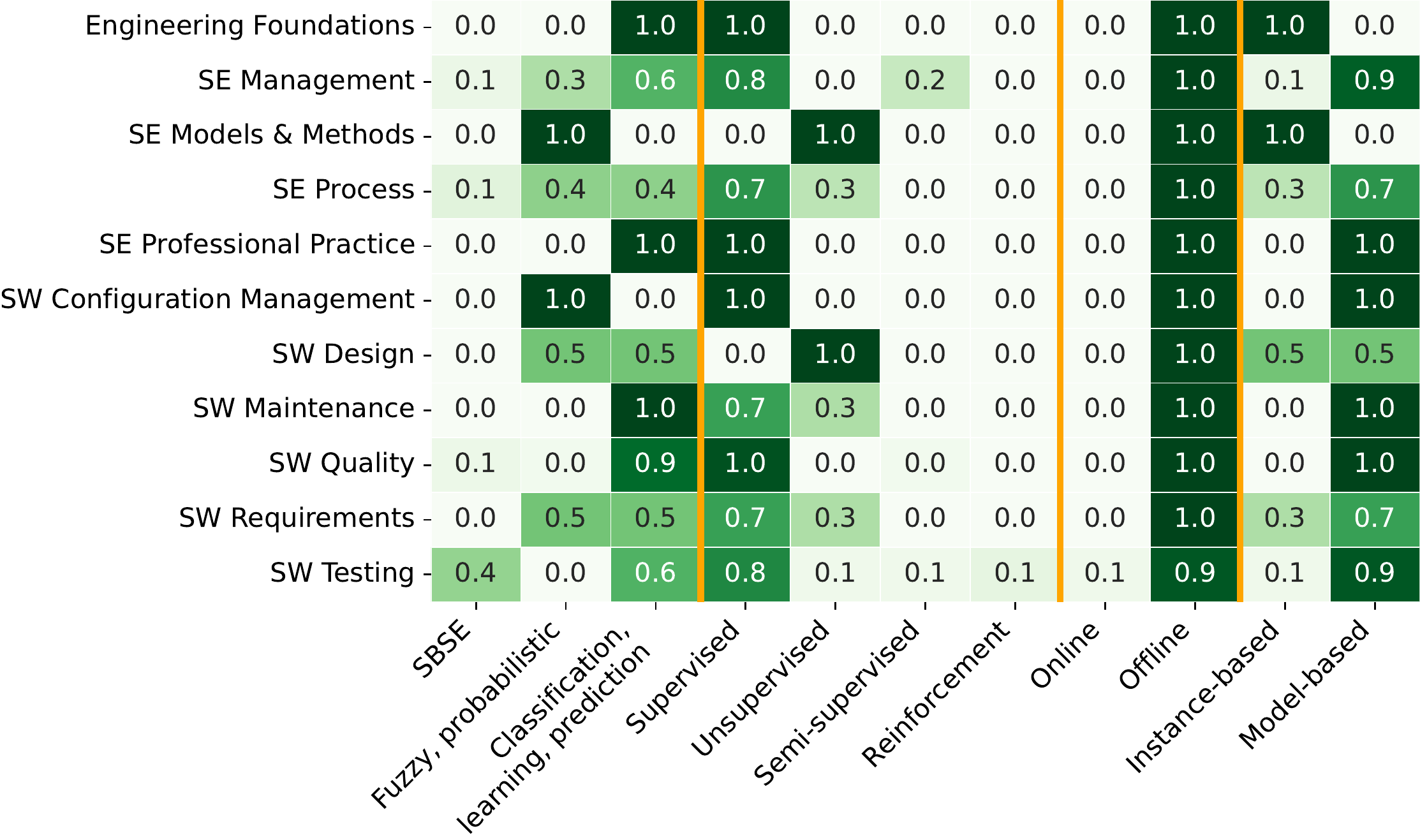}
  \caption{Percentage distribution of publications across ML techniques and SWEBOK KAs.
  (The ML axes are separated by vertical lines.)}
  \label{fig:techniques}
\end{figure}

In Fig.~\ref{fig:techniques} we visualize the percentage distribution of the studies
across their assigned SWEBOK KAs (Table~\ref{tab:RQ1tab}) and
ML techniques (Table~\ref{tab:MLclass}).
The four axes are separated by vertical lines, and
for each KA,
the techniques of each axis sum up to one.
For instance,
10\% of studies assigned to \emph{SE Management}
employ computational search and optimisation techniques,
30\% employ fuzzy and probabilistic methods, and
60\% employ classification, learning, and prediction techniques.
As expected from Table~\ref{tab:MLclass}, in all KAs
researchers mainly employ classification, learning, and prediction techniques, and
apply supervised, batch/offline, model-based learning.
Interestingly,
in the areas of \emph{SE Process}, \emph{Software Design}, and
\emph{Software Requirements},
a considerable number of studies involve fuzzy and probabilistic methods
of unsupervised, instance-based learning.
In \emph{SE Testing} we also observe a lot of reviews
on computational search and optimisation techniques
(the field known as Search Based Software Engineering---SBSE).

\textbf{Classifying the manually extracted ML techniques
(Section~\ref{sec:data-extraction})
according to the SE tasks outlined in Section~\ref{sec:rq1}
results in no insight---the same algorithms appear to be used in all SE tasks.}
What differentiates their use is the ML application task.
In ML4SE, we recognize the following ML tasks:
classification, clustering, regression;
pattern discovery;
dimensionality reduction;
information retrieval;
stochastic search;
generation.
We classified the identified ML techniques according to these ML tasks, and
also included two additional categories: hybrid and miscellaneous.
Hybrid techniques concern combinations of the aforementioned ML tasks
(\eg stochastic search and clustering),
while miscellaneous include techniques that do not fall into a
particular category.
In Table~\ref{tab:techniques}
we list the techniques used in each ML task.

\begin{table}[t]
  \caption{ML Techniques Grouped by Application Task}
  \label{tab:techniques}
  \centering
  \resizebox{\textwidth}{!}{
  \begin{tabular}{p{20cm}}
    \textbf{Classification, Clustering, Regression:}
      Artificial Neural Network
      (Back Propagation, Multi-Layer Perceptron, Cascade Feed Forward,
      General Regression, Radial Basis Function, Convolutional---CNN,
      Probabilistic, Graph, Recurrent---RNN, Long Short-term Memory---LSTM,
      Gated Recurrent Unit---GRU, Fuzzy, Siamese, Deep Belief,
      Restricted Boltzmann Machine, Generative Adversarial,
      Autoencoder, Encoder-Decoder);
      Bayesian Network
      (Meta-learner, Converging Star, Causal, Dynamic,
      Transfer, Weighted, Augmented, Boosting);
      Binary Classifier;
      Case-based Reasoning;
      Clustering variations
      (Hierarchical Agglomerative,
      Hierarchical Conceptual with COB-WEB,
      Incremental Diffusive,
      Self Organizing Map,
      LIMBO-based Fuzzy Hierarchical);
      Decision Tree
      (C4.5, C5.0, M5, Partial, J48, Alternating, Naive Bayes,
      Classification and Regression,
      Reduced Error Pruning Tree, Tree Discretiser,
      Chi-square Automatic Interaction Detection,
      Plausible Justification Tree);
      Ensemble Learner (Bagging, Logit Boost, AdaBoost, XGBoost, Analogy-based);
      Expectation-Maximization algorithm;
      Feature-gathering Dependency-based Software Clustering (SArF);
      Fisher's Linear Discriminant;
      Fuzzification/Defuzzification;
      Gaussian Mixture Model;
      Grid Search;
      Instance-based Learner (k-Nearest Neighbors, K-star, IBk);
      K-Means, Fuzzy K-Means, Fuzzy C-Means, X-Means, K-Medoids,
      SPK-Means, Affinity Propagation;
      Learning Finite Automata (\eg hW-Inference);
      Most Common Attribute;
      Naive Bayes Classifier
      (Binarized, Bernoulli, Multinomial, Tree Augmented, Gaussian);
      Q-Learning;
      Random Forest
      (Adaptive Deep Forest, Rotation Forest,
      Isolation Forest, Global Abnormal Forest);
      Regression Analysis
      (Univariate Linear, Multivariate,
      Multivariate Adaptive Regression Splines);
      Regression variations
      (Linear, Meta-learner Linear, Ensemble Linear,
      Multiple Linear with Backward Elimination/Stepwise Selection,
      Logistic, Additive, Ordinary Least Square, Symbolic, Polynomial,
      Multiplicative Adaptive Spline, Projection Pursuit);
      Support Vector Machine (SVM), Relevance Vector Machine, One-class SVM;
      Systematic Method for Software Architecture Recovery (SysMar)

    \textbf{Pattern Discovery:}
      Association Rule Learning (\eg Apriori, FP-Growth, ECLAT);
      Non-Nested Generalisation (NNGE);
      Repeated Incremental Pruning to Produce Error Reduction (Ripper);
      Zero Rule, One Rule, Fuzzy Rule

    \textbf{Dimensionality Reduction:}
      Correlation Feature-based Selection;
      Isomap;
      Principal Component Analysis;
      Self-organizing Map, Generative Topographic Map;
      Singular Value Decomposition

    \textbf{Information Retrieval:}
      Artificial Immune System, Artificial Immune Recognition System;
      Best Match 25;
      Binary Independence Model;
      Contrastive Analysis;
      Language Model;
      Latent Dirichlet Allocation
      (LDA---Delta, Dynamic, Labeled, jsLDA, Maximum-likelihood Representation,
      with Gibbs Sampling, Temporal);
      Latent Semantic Analysis;
      Latent Semantic Indexing (LSI), Probabilistic LSI;
      Probabilistic Inference Network;
      Topic Model
      (Correlated, Relational, Biterm, Multi-feature, Citation Influence, Collaborative);
      Vector Space Model (VSM), Generative VSM

    \textbf{Stochastic Search:}
      Ant Colony Optimization, Ant Colony-based Data Miner (Ant-Miner);
      Bat Algorithm;
      Cuckoo Search;
      Firefly Algorithm;
      Gene Expression Programming;
      Genetic Algorithm;
      Genetic Programming;
      Group Search Optimization;
      Hill Climbing;
      Particle Swarm Optimization;
      Sequential Minimal Optimization;
      Simulated Annealing;
      Tabu Search

    \textbf{Generation:}
      Domain-specific Language Guided Model;
      Memory-augmented Neural Network
      (Memory Network, End-to-End Memory Network, Recurrent Entity Network);
      \emph{n}-gram Language Model;
      Non-recurrent Neural Network
      (CNN, Dynamic CNN, Convolutional Sequence to Sequence Learning,
      BERT, Transformer, WaveNet);
      Pre-trained Word Embeddings
      (Word2Vec, FastText, Global Vectors for Word Representation---GloVe,
      Contextualized Word Vectors---CoVe, Character to Word---C2W);
      Probabilistic Grammar
      (Abstract Syntax Tree, Suffix Tree, Context-Free Grammar,
      Probabilistic Context-Free Grammar, Tree Substitution Grammar,
      Tree Adjoining Grammar);
      Recurrent Neural Network
      (Vanilla RNN, LSTM, GRU, Fast-Slow RNN, Recurrent Highway Network)

    \textbf{Hybrid:}
      Artificial Neural Network-Evolutionary Programming;
      Evolutionary Decision Tree (LEGAL Tree);
      Genetic Algorithm-Artificial Neural Network;
      Genetic Algorithm-Support Vector Machine
      with a linear/radial basis kernel function;
      Genetic Programming-Decision Tree;
      Hyper-heuristic Evolutionary Algorithm-Decision Tree (HEAD Tree);
      Particle Swarm Optimization-Artificial Neural Network;
      Simulated Annealing-Probabilistic Neural Network

    \textbf{Miscellaneous:}
      CODEP (COmbined DEfect Predictor) algorithm
      for cross project defect prediction~\cite{APSW19};
      ML model for system-level test case prioritization using
      black-box metadata and natural language test case descriptions~\cite{CGB20};
      SwiftHand tool for automated Android GUI testing~\cite{ZMI17}
  \end{tabular}}
\end{table}

\section{Discussion and Implications}
\label{sec:discussion}
In the past twenty years,
diverse studies
identified, summarized, and assessed the contribution of ML in SE.
Despite the considerable number of ML-based approaches that have been proposed
in the associated academic literature \textbf{for a large number of SE tasks},
their practical adoption and application
by the industrial community appears limited.
From the analysis of RQ2 (Section~\ref{sec:rq2}) we conclude that
some reasons for this could be the lack of:
empirical work evaluating the quality and cost-effectiveness of these approaches,
comparative analyses contrasting their performance and execution time
to that of conventional statistical techniques, and
industrial trials assessing their relevance, efficiency, and scalability
in the context of large projects.
A starting point to address the latter and
improve the quality of future ML4SE studies could be
thorough industrial case studies
aiming to unveil any obstacles in the adoption of ML approaches,
as suggested by Wen \etal~\cite{WLLH12}.

\begin{implication}
Further empirical validation studies, comparative analyses, and industrial trials
assessing and contrasting the
quality
of the proposed ML techniques for SE tasks to
that of conventional statistical approaches
are required to amplify the salience of the former,
and boost their practical adoption and application.
These are mostly needed in the areas of
software quality, testing, and engineering management.
\end{implication}

Apart from the lack of empirical and comparative analyses
assessing the quality of existing ML models,
another obstacle in their adoption and advancement could be the lack of
up-to-date methods for developing reliable and selecting suitable models.
This deficiency was mainly observed in the tasks of
smell detection,
software fault and maintainability prediction, and
requirements volatility prediction.
Particularly in smell detection,
the inconsistency observed by Sharma and Spinellis~\cite{SS18}
between smell definitions and detection methods seems to be a result of
the absence of smell literature establishing standards for
smell definitions,
their implementations, and
commonly used metrics.
Similar remarks are noted in software maintainability prediction
concerning maintainability definitions and characteristics~\cite{ECIA20}.
ML models are typically founded on clearly defined principles and rules
extracted from the training data
(otherwise learning would simply correspond to memorization)~\cite{MB18}.
Therefore,
it seems that the growth of ML models is indirectly impacted
by inadequately established SE concepts and
general literature shortcomings.

\begin{implication}
The academic research community could consider taking a step back
to address its fundamental SE literature shortcomings and inconsistencies
including outdated and deficient methods,
which appear to impact the development and selection of
robust and reliable ML models for SE tasks.
Such deficiencies are mainly observed in the tasks of
smell detection,
software fault and maintainability prediction, and
requirements volatility prediction.
\end{implication}

Looking at Table~\ref{tab:RQ1tab} it appears that SE tasks
falling into more technical SWEBOK KAs such as software quality and testing
are more frequently addressed with ML approaches,
compared to KAs targeting the human factor,
such as SE professional practice or software requirements.
This could be a result of the inadequate performance of
existing ML models of the latter KAs,
as observed in the results of RQ2 (Section~\ref{sec:rq2}),
discouraging developers from performing further studies in these KAs,
and prompting them to more fertile grounds (\ie technical KAs)---a
tendency known as the \emph{streetlight effect}~\cite{BA15, Bas21}.
It could be the case that human-centered SE tasks entailing a high rate
of subjectivity are more difficult to be approximated with ML,
either due to inherent limitations imposed by subjectivity
on the evaluation of the respective ML models,
or the lack of ground truth datasets.
Interestingly,
a similar distribution of the KAs is also observed
in the reverse field of SE for AI,
according to a recent survey by Mart{\'{\i}}nez{-}Fern{\'{a}}ndez \etal~\cite{MBFO22}.

Regarding data availability,
in a recent study about Mining Software Repositories data papers~\cite{KKDS20}
the identified datasets about developers' attributes
were considerably fewer than those about software attributes,
such as version control system data, or software faults, failures, and smells.
Gathering and labeling data about developers' mental, behavioral, or
sentimental characteristics has long been a challenge for the research community,
remaining an open issue~\cite{KKDS20}.
Similarly,
automatically extracting software requirements
from natural language documents with ML,
as suggested by Bakar \etal~\cite{BKS15},
has been difficult, because the usually unstructured nature of the documents
impedes their systematic processing.
On the contrary,
gathering data from a program's execution, testing, and debugging phases
can be automatically achieved with various tools. The preponderance of
datasets with software faults, failures, and smells
in our aforementioned study~\cite{KKDS20} confirms this.
Review authors suggest the incorporation
of developers' real-time emotion-sensing biometric data
(\eg heart rate, eye blinking, and electroencephalogram)
in human-centered datasets
to potentially improve a ML-based system's performance
(\eg for emotion detection)~\cite{GFSF19}.
However,
disclosing and employing such private information
might raise privacy concerns~\cite{RASV14}.

\begin{implication}
SE tasks associated with human-centered SWEBOK KAs
such as SE professional practice or software requirements appear less tackled
with ML techniques
compared to more technical ones including software quality and testing.
Addressing the weak spots requires investment into the difficult tasks of
collecting and labeling training data about developers' attributes, and
evaluating the typically subjective accuracy of the ML-based systems.
\end{implication}

Complementary to the aforementioned difficulties of
collecting and labeling data for human-centered SE tasks,
further concerns related to the datasets' quality were remarked
in the secondary studies.
Data quality is often dubious due to the inadequate documentation of
the data collection,
cleaning,
and pre-processing methods (\ie the data pipeline)~\cite{HBBG12}.
These shortcomings are not only encountered in ML4SE;
they further expand to SE for AI~\cite{MBFO22},
suggesting major field-agnostic data issues that need to be resolved.
Along with better documentation,
the automation of the data pipeline activities through ML frameworks
such as Auto-WEKA~\cite{THHL13} or Auto-sklearn~\cite{FKES19}
could further improve data quality, and consequently,
the performance of the trained ML models~\cite{Que19}.

\begin{implication}
To increase the quality trustworthiness of training datasets,
researchers should document their data collection and pipeline processes, and
investigate the cost-benefit of automating them through ML frameworks.
\end{implication}

Additional dataset issues arise from their industrial relevance and scalability.
Some practitioners seem reluctant to utilize ML models from academia
that are only trained on open data,
considering them less realistic and representative
of industrial contexts~\cite{HTG19}, and
less scalable to large projects~\cite{BRA14}.
In addition,
they may also fear that ML models trained solely on publicly available data
will not produce novel results,
or be unwilling to work with ML models developed by a different
organization---also called the \emph{not invented here syndrome}~\cite{PD15}.
These last two cases were also observed by us~\cite{KKDS20}
with regard to the use of data papers,
recommending to methodology researchers, conference program committees, and
journal editorial boards
the embracement of a procedure similar to that of
pre-registered studies~\cite{HI18}
(\ie publishing a data paper and then employing it
for empirical SE research).
To strengthen the industrial application of ML models to SE tasks,
the same paradigm could also be employed in the ML and SE intersection
by promoting the advance publication of datasets used
in the development of ML-based systems.
Furthermore,
industrial partners should consider
sharing more of their proprietary data
to help academia build more robust and realistic ML models,
which will likely benefit the industry as well.

\begin{implication}
To improve the industrial relevance, scalability, and performance of ML models,
practitioners might want to consider
sharing more of their proprietary data with academia.
Moreover,
methodology researchers, conference program committees,
and journal editorial boards could investigate the value of
adopting a research paradigm where training datasets are published
before the ML models that use them.
\end{implication}

In respect to the incrementality of employed ML techniques in SE
(Section~\ref{sec:rq3}),
there seems to be a vast preference for batch/offline learning,
despite the benefits of online learning.
Online learning is more computationally effective for dealing with new data,
and also works well for systems that receive data as a continuous flow and
need to adapt to change rapidly or autonomously~\cite{KJ18}.
Motivated by this,
there has been an emergence of promising incremental data retrieval methods
and tools recently (\eg the works by
Mastorakis \etal~\cite{MGAZ18},
Fu \etal~\cite{FLWL19},
Aydin \etal~\cite{AA17}).
At the same time,
other science and engineering disciplines, such as
healthcare~\cite{PKKE20} and transportation engineering~\cite{NNBA19},
have already started experimenting with online/incremental ML techniques.
For instance,
in the earthquake engineering domain,
there is an active research area concerned with
developing ML-based structural control schemes
for earthquake mitigation~\cite{XSPD20}.
In this context,
Suresh \etal~\cite{SNNS10} accomplished real-time online adaptation
of an artificial neural network-based controller
through the use of an extended minimal resource allocation network.
The authors argue that
online ML-based structural controllers appear more effective
in mitigating the adverse effects of earthquake hazards on buildings
than traditional approaches.

The SE community could be inspired by cross-domain online ML applications
like the aforementioned,
and adapt them to the SE field, \eg for real-time monitoring of applications.
Promising data sources include
build and execution logs,
crash reports,
security incidents,
integrated development environment and user interactions,
telemetry, and
the internal usage of code abstractions (\eg software functions, API endpoints).
An instance of online ML application in a middleware could involve
the real-time verification of the availability of the associated endpoint servers
as well as
the real-time analysis of potentially defective software components
to determine potential self-healing actions.

\begin{implication}
Online and incremental ML-based applications in SE provide a fertile ground
for further research,
due to their computational effectiveness, rapid changeability and adaptability, and
thanks to recent advances in incremental data retrieval methods and tools.
\end{implication}

The general idea of experimenting with ML approaches applied
in different domains and contexts
is also supported by the authors of some secondary studies
(Section~\ref{sec:rq2}).
Recommendations pertain to
combining probabilistic or search-based techniques
with ML approaches~\cite{BRA14, MKR17, AZAB17, HTG19, IHA16}, and
transferring novel methods from relevant fields~\cite{IAA15, WLLH12}.
The lower percentage of studies classified in the categories of
\emph{Computational search and optimization techniques} (14\%) and
\emph{Fuzzy and probabilistic methods for reasoning in the presence of uncertainty} (20\%) as well as
the limited number of employed hybrid ML techniques,
compared to the ones under the category of
\emph{Classification, Clustering, Regression} (see Section~\ref{sec:rq3}),
aligns with the above recommendations,
suggesting room for improvement.
Furthermore,
the encouraging achieved results in effort estimation and defect prediction
with hybrid models,
as concluded by Malhotra \etal~\cite{MKR17},
is a positive indicator for performing more experiments with them.

\begin{implication}
Hybrid ML techniques encompassing probabilistic or search-based approaches, and
cross-disciplinary novel ML methods
have yielded positive results in certain SE tasks,
hence they might be worth of further investigation.
\end{implication}

\section{Threats to Validity}
\label{sec:threats}
We use the classification scheme proposed by Ampatzoglou \etal~\cite{ABAV19}
to classify the limitations of this study.
This is inspired by the \emph{planning} phase of reviews
(\ie search process, study selection, data extraction, and data analysis)~\cite{KC07},
and is extended with an additional category that concerns threats
from the entire lifecycle of the review~\cite{ABAV19}.

\par{\textbf{Study Selection Validity}}
The adopted study search and selection strategies are associated
with the risk of missing relevant studies.
Some research may have been missed as a result of
the selected year range in the automated search (\ie 2015--2020),
the preferred digital libraries, and
the constructed search strings (Section~\ref{sec:search}),
or the applied selection criteria (Section~\ref{sec:ic_ec}).
The year 2015 was considered the inflection point
for the joint evolution of the two fields,
as elaborated in Section~\ref{sec:automated-search},
allowing us to center our analysis on the interdisciplinary growth.
To reduce the threat of missing relevant reviews,
we searched the digital libraries
that are most likely to include the majority of studies in ML4SE.
However,
we cannot eliminate the chance
that we may have missed some germane studies
while conducting the automated search in these databases.
To cope with the interdisciplinarity of the subject area,
the keywords for our search strings (Table~\ref{tab:searchkeys})
were derived from established sources.

The quality assessment process (Section~\ref{sec:qa}) is also associated
with a few threats.
A number of pertinent studies were deliberately excluded
to ensure high quality of our study results,
as recommended in the adopted guidelines~\cite{KC07}.
Furthermore,
the DARE-4 framework employed in the quality evaluation
of the reviews does not cover all quality facets~\cite{CFFQ21}.
This is a common threat of all existing quality assessment frameworks,
and it is a recommended practice for tertiary reviews
to select the framework that best satisfies the research goals~\cite{CFFQ21}.
For this reason we selected DARE-4,
which was deemed the most appropriate framework for our evaluation,
and is also the most commonly used one in SE tertiary studies~\cite{CFFQ21}.

\par{\textbf{Data Validity}}
One potential limitation stems
from the data extraction process (Section~\ref{sec:data-extraction}).
Some secondary studies did not provide all the information needed,
and we had to infer it.
For instance,
some studies did not include a summary list of the associated primary research
(\eg~\cite{LLN20}).
To extract the number of primary studies and their publication years
in these cases (Tables~\ref{tab:dataextractiontab},~\ref{tab:dataextractiontab2}),
we looked up the bibliography section,
omitting irrelevant research referenced, \eg in introduction or related work.
In addition,
some secondary reviews did not cite the employed research method
(\eg~\cite{TAB19, SS18}),
despite their detailed descriptions.
For these,
we considered the method description and the complete review structure
to infer the adopted guidelines.
For example,
SLRs following Kitchenham and Charter's guidelines~\cite{KC07}
typically report their
research questions,
search process,
study selection method,
quality assessment, and
data extraction process.

Another data validity threat arises from one of the composing axes
of the ML classification scheme (Table~\ref{tab:MLclass}).
Specifically,
the \emph{role of AI in SE} applies to a broader field than ML (\ie AI).
Consequently, it could be considered less appropriate for the categorization
of studies targeting the application of ML in SE.
We decided to use this axis because we considered that
it would complement the results with additional useful information.
(To the best of our knowledge,
there is no available study characterizing and categorizing the role of ML in SE.)
To this end,
all studies were successfully assigned to a category,
while the data extractor and data checker maintained high inter-rater reliability,
suggesting that
the categories of this axis are suitable for the categorization
of ML4SE research.
Follow-up studies could validate the extent of congruence
between the role of AI in SE and that of ML in SE.

\par{\textbf{Research Validity}}
The study's research validity is partially concerned with the extent to which
the results of our tertiary review can be generalized
to the subject population.
Therefore,
one potential issue stems from assessing
whether the secondary studies are representative of all the relevant studies
in the subject area.
To minimize this threat,
we performed a comprehensive multi-phase search procedure
(automated, manual, backward, and forward snowballing search)
in more than one digital libraries (Section~\ref{sec:search}),
during which we tried to be as inclusive as possible
with respect to the selection criteria (Section~\ref{sec:ic_ec}),
following established guidelines~\cite{KC07}.

Other major threats to the research validity stem from
the steps during which we followed manual processes involving subjective judgment.
These include the
manual and backward snowballing search processes (Section~\ref{sec:search}),
the study selection (Section~\ref{sec:selection-process}) and
quality assessment (Section~\ref{sec:qa}),
the data extraction process,
the classification of the studies using the SWEBOK KAs and the multi-axis ML scheme,
the extraction of the tackled SE tasks by ML using the open coding practice,
the identification of implications for further research in ML4SE,
and the detection of the employed ML techniques in SE
(Section~\ref{sec:data-extraction}).
The reliability of these processes was improved
by engaging multiple raters,
and by basing them on standard research methods.
However,
we recognize that validity threats stemming from manual processes
entailing subjective judgment cannot be nullified~\cite{PVK15}.

\section{Conclusion and Recommendations}
\label{sec:conclusion}
In our tertiary review we systematically retrieved 140 secondary studies
in ML4SE,
and analyzed 83 of them that satisfied a set of recommended quality criteria.
These 83 reviews span the years 2009--2022,
were authored by 274 researchers affiliated with 140 institutions, and
entail 6\,117 primary works published between 1990--2021.
To analyze the reviews we followed established guidelines
and designed a protocol that was internally agreed by all authors.
The analysis was performed by hand and consisted of:
the classification of the reviews using the SWEBOK KAs and subareas;
the extraction of SE tasks tackled with ML from the reviews;
the extraction of SE topics for further research using ML from the reviews;
the categorization of the reviews
using a four-axis ML classification scheme that was synthesized from two sources; and
the extraction of the ML techniques employed in the reviews.
Through these manual processes the following key findings were obtained.

\begin{itemize}[leftmargin=*]
  \item The majority of secondary reviews in ML4SE target the SWEBOK KAs of
  software quality and testing, and SE process.
  Human-centered KAs such as SE professional practice and software requirements appear
  less tackled with ML techniques,
  due to the subjectivity entailed in the evaluation of the models,
  and the difficulty of collecting and labeling training data
  about developers' characteristics.
  \item With regard to the role of AI in SE,
  most studies pertain to \emph{Classification, learning and prediction} tasks,
  and apply supervised learning.
  In terms of generalizability,
  model-based learning is vastly preferred.
  Despite the demonstrated benefits of online/incremental learning and
  the emergence of relevant tools,
  batch/offline learning is overwhelmingly used.
  \item Some major obstacles to the advancement of ML techniques result from
  the training datasets and SE literature discrepancies.
  Validity issues often arise from
  undocumented data collection and non-automated data pipeline processes,
  while the absence of proprietary data burdens the industrial relevance,
  scalability, and performance of ML models.
  Outdated and deficient methods obstruct researchers from developing robust and
  selecting appropriate ML models for SE tasks.
\end{itemize}

\par{\textbf{Recommendations for Researchers}}
From Section~\ref{sec:rq2-general} we summarize the following suggestions.
Researchers should further assess the proposed ML techniques,
compare them to conventional statistical approaches, and
evaluate their scalability, performance, and cost-effectiveness
in industrial settings as well as
through further empirical analyses.
To this end,
more in-depth case studies with practitioners should be conducted.
A starting point to improve the quality of ML models is by
optimizing their hyper-parameters,
addressing class imbalance in training datasets, and
developing concrete methods for building reliable systems.
Hybrid, ensemble, and incremental ML techniques, and
cross-domain methods comprise promising areas for additional experimentation.

\par{\textbf{Recommendations for Practitioners}}
Practitioners can benefit from existing ML4SE research
through the various published ML-based open source tools for SE tasks.
To select the most applicable ones,
they can consult the associated meta-analyses
summarized in this study~\cite{HTG19,ECIA20,MDDC19,APSW19,SB20}.
ML-based tools can either be applied directly to their corporate projects, or
be used as baselines to compare the performance of their own tools.
In both cases
care should be taken regarding the false-positiveness, fine-tuning, and training
of the adopted systems.
Furthermore,
practitioners can take advantage of the diverse ML4SE research implications
to improve their existing ML systems,
or produce new ones to satisfy further needs.

To help researchers build more accurate ML models,
the industry needs to release more open-source, large-scale datasets, and
collaborate with academia in industrial trials and case studies.
Collaborations can happen through
funded research projects,
internships,
regular workshops and seminars,
conference participation,
technology transfer test labs (for piloting research ideas), and
the involvement of industry partners in research education~\cite{GPO16}.
Through practitioners' feedback and support
researchers will be able to apply their models in large projects,
understand the industry's needs, and
improve their methods.
Closing this loop should provide practitioners with better ML-based SE tools.

This is the first systematic tertiary study providing a
comprehensive overview of the current state of the practice in ML4SE.
With these final considerations
we hope to increase awareness on certain issues identified in the intersection
of the two fields,
and steer researchers' attention towards under-explored areas and topics
requiring further investigation.

\begin{acks}
This work has received funding from
the European Union's Horizon 2020 research and innovation programme
under grant agreement No. 825328 (FASTEN project).
\end{acks}

\bibliographystyle{ACM-Reference-Format}
\bibliography{ml4se,qa-studies}

\end{document}